\documentclass[letterpaper, 10 pt, journal, twoside]{IEEEtran}
\usepackage{graphicx}
\usepackage{epsfig}
\usepackage{times}
\usepackage{amsmath}
\usepackage{amsthm}
\usepackage{thmtools,thm-restate}
\usepackage{amssymb}
\usepackage[section]{placeins} 
\usepackage{placeins}
\usepackage{amsmath} 
\usepackage{multirow}
\usepackage{graphicx}
\usepackage[normalem]{ulem}
\usepackage{subcaption}
\usepackage{algorithm,tabularx}
\usepackage{algpseudocode}
\usepackage{amsfonts}
\usepackage{cases}
\usepackage{romannum}
\usepackage{textcomp}
\usepackage{xcolor}%
\usepackage{textcomp}
\providecommand{\U}[1]{\protect\rule{.1in}{.1in}}
\providecommand{\U}[1]{\protect\rule{.1in}{.1in}}

\IEEEoverridecommandlockouts
\newtheorem{assumption}{Assumption}
\newtheorem{theorem}{Theorem}
\newtheorem{corollary}{Corollary}
\newtheorem{lemma}{Lemma}

\newtheorem{remark}{Remark}
\newtheorem{definition}{Definition}

\useunder{\uline}{\ul}{}
\makeatletter
\newcommand{\multiline}[1]{  \begin{tabularx}{\dimexpr\linewidth-\ALG@thistlm}[t]{@{}X@{}}
#1
\end{tabularx}
}
\makeatother
\usepackage{cite}

\usepackage{picins}
\usepackage{centernot}

\usepackage{enumitem}
\setlist[itemize]{leftmargin=*}
\usepackage{makecell}
\usepackage[normalem]{ulem}
\usepackage{cuted}
\usepackage{hyperref}
\usepackage{stfloats}
\usepackage{lscape}
\usepackage{algorithm}
\usepackage{algpseudocode}

\newcommand{\tsup}[1]{\textsuperscript{#1}}
\newcommand{\T}{\top}
\newcommand{\I}{\mathbf{I}}
\newcommand{\0}{\mathbf{0}}
\newcommand{\diag}{\text{diag}}
\newcommand{\bmtx}[1]{\begin{bmatrix}#1\end{bmatrix}}

\DeclareMathOperator{\rank}{rank}

\renewcommand{\u}{\bm{u}}

\IEEEoverridecommandlockouts
\begin{document}

\title{\bf\Large Funnel-Based Online Recovery Control for Nonlinear Systems With Unknown Dynamics}

\author{Zihao Song, Shirantha Welikala, Panos J. Antsaklis and Hai Lin\thanks{This work was supported by the National Science Foundation under Grant IIS-2007949. Zihao Song, Panos J. Antsaklis, and Hai Lin are with the Department of Electrical Engineering, University of Notre Dame, Notre Dame, IN, USA. e-mail: \{{\tt zsong2, pantsakl, hlin1}\}{\tt @nd.edu}. Shirantha Welikala is with the Department of Electrical and Computer Engineering, Stevens Institute of Technology, Hoboken, NJ, USA. e-mail: {\tt swelikal@stevens.edu}.
% The first author sincerely appreciates Dr. Shirantha Welikala for several fruitful discussions.
}}
\maketitle
\thispagestyle{empty}

\begin{abstract}
   In this paper, we focus on recovery control of nonlinear systems from attacks or failures.
   The main challenges of this problem lie in (1) learning the unknown dynamics caused by attacks or failures with formal guarantees, and (2) finding the invariant set of states to formally ensure the state deviations allowed from the nominal trajectory.
   To solve this problem, we propose to apply the Recurrent Equilibrium Networks (RENs) to learn the unknown dynamics using the data from the real-time system states. The input-output property of this REN model is guaranteed by incremental integral quadratic constraints (IQCs). Then, we propose a funnel-based control method to achieve system recovery from the deviated states. In particular, a sufficient condition for nominal trajectory stabilization is derived together with the invariant funnels along the nominal trajectory. 
   Eventually, the effectiveness of our proposed control method is illustrated by a simulation example of a DC microgrid control application.
\end{abstract}

%-------------------------------------------------------------
% \vspace{-2mm}
\section{Introduction}\label{sec:intro}
% \vspace{-1mm}

% 1. System resilience $-->$ recovery

\subsection{Background and motivations}

Resilience refers to the capability of a natural or engineered system to withstand disturbances, attacks, or failures, recover its functionality with graceful degradation, and adapt to new operating conditions.  Resilience becomes a focal point in the design of next-generation complex engineering systems, such as autonomous vehicles, robotic networks, and smart grids, where the systems are open to uncertain dynamic environments and are vulnerable to cyber-physical attacks. 

Most of the existing work focuses on fault detection \cite{wang2017improved}, fault tolerant control \cite{abbaspour2020survey}, or security enhancement \cite{wan2023brief}. However, the study of recovery control is still relatively lacking. By recovery control, we mean designing a control mechanism that not only maintains system stability after faults or disturbances, but also actively drives the system states back to their desired operating points or equilibrium conditions after the occurrence of a fault, attack, or disruption.

Recovery control poses significant challenges. First, we need to ensure formal performance during the transition phase, i.e., the sets of allowable state deviations with respect to the nominal trajectory, so that the systems' stability and safety are not violated while the system is recovering from faults or attacks.
Secondly, system recovery should usually be conducted online, as system performance will be continuously degraded if not properly and quickly handled, which can cause secondary damage \cite{wang2024comprehensive}. 
Thirdly, recovery control must deal with unknown dynamics as faults, attacks, and topology changes often alter the system dynamics, which may make conventional control schemes difficult to apply.
In this paper, we are, therefore, motivated to develop an online funnel-based control framework for the recovery control of systems with unknown dynamics.

% {\color{red} Most of existing work ... fault detection, fault tolerant... security ... However, the study of recovery control is still lacking... By recovery control, we mean ...   return the states of the systems towards their desired operating points... }
% {\color{gray}
% While other resilience tasks focus on detection, isolation or the resistance and tolerance of the impacts of the attacks or failures, system recovery emphasizes on reconfiguration, adaptation, and returning the system to stable/nominal operating points after disruption.}

% {\color{red} Recovery control poses significant challenges. First, we need to ensure the formal performance during the transition phase. Secondly,} system recovery should usually be conducted online, since system performance will be continuously degraded if not properly and quickly handled, which may cause secondary damages \cite{wang2024comprehensive}.
% {\color{red} Thirdly, ... unkown dynamics ...
% We are therefore motivated to ... recovery control... }

% 2. Existing methods for recovery control

%-------------------------------------------------------------
\subsection{Related work}
% {\color{red} Shall we move this subsection out as a new section "Related work"?}

From the control theoretical perspective, early efforts for system recovery control mainly focused on robust control approaches, such as $H_{\infty}$ control \cite{saeki1992h,weng1998h}, fault-tolerant control (FTC) \cite{jiang2012fault} and invariant-set theory \cite{yu2018control,mccloy2019set}, which ensure stability/performance margins under attacks or faults.
Although these methods can provide strong guarantees, they usually rely on accurate modeling assumptions with the knowledge of structured unknown dynamics. 
Besides, these methods are mostly designed offline to withstand attacks or failures but cannot proactively and efficiently handle their impacts. 
There have also been studies on resilient consensus \cite{cheng2024resilient,yuan2024resilient} and safety-guaranteed designs (e.g., control barrier functions (CBFs) \cite{cavorsi2023multirobot} and barrier Lyapunov functions (BLFs) \cite{romdlony2016stabilization}), but the adaptability to substantial changes in the system dynamics is still limited.

Recent years have seen the development in this domain including reconfiguration-based methods (e.g., active FTC \cite{abbaspour2020survey,cai2025active}, and control reallocation \cite{cai2025active,ergoccmen2025reconfigurable}), adaptation-based methods (e.g., adaptive control \cite{liu2025adaptive}, switching-based control \cite{shirazi2019robust,ke2023uniform} and adaptive model predictive control (AMPC) \cite{li2025cyber}), learning-based methods (e.g., reinforcement learning (RL) \cite{lang2022deep}, data-driven methods \cite{van2021matrix,xu2025data} and Gaussian process (GP)-based methods \cite{ma2024adaptive}) and game theoretic methods \cite{gautam2021cooperative,gill2024systematic}.
However, the aforementioned methods may not be directly applicable to scenarios where there exist unknown dynamics in the systems. 
If not properly handled, these dynamics may deteriorate system performance or even result in instability of the systems.

Apart from the unknown dynamics, finding the formal guarantees for the allowable state deviations with respect to the nominal trajectory is generally non-trivial.
% {\color{gray} Besides, finding the formal guarantees for the allowable state deviations with respect to the nominal trajectory is generally non-trivial. 
% Another main concern in recovery control is the unknown dynamics caused by attacks or failures, which are usually time-varying and bounded.} 
Even though methods like tube-based MPC \cite{hill2023tube} can ensure the invariance of system states around the nominal trajectory under unknown dynamics, they may not be tractable for complex nonlinear systems and lack formal guarantees for stability and transition performance during system recovery. 
Compared to the aforementioned methods, funnel-based control \cite{reynolds2021funnel,kim2023optimization}, as a relatively new method for path planning, appears to be specifically suitable for system recovery control, since multiple constraints such as convergence behaviors, input saturation, and safety constraints can be combined in the solutions and invariant funnels can be systematically derived that allow maximal state deviations with respect to nominal trajectories. However, unknown dynamics are not involved in existing funnel-based control designs \cite{reynolds2021funnel,kim2023optimization}.

%-------------------------------------------------------------
\subsection{Our approach and contributions}

% {\color{red} To handle unknown dynamics, we ... learning-based approaches ... }

% {\color{gray} In this paper, we consider the problem of online recovery control for systems with unknown dynamics.  To handle these unknown dynamics, existing work in recent years mainly focuses on adaptive control \cite{wu2024performance,ma2024adaptive}, data-driven control \cite{van2021matrix,xu2025data}, and learning-based methods \cite{shafa2025reachable}. Among these approaches, we are, in particular, interested in using learning-based methods, since they are independent of system dynamics, and can be adaptable to system changes, which is well-suited for recovery control and can be easily combined with the proposed funnel-based method.} 

% In this paper, we consider the problem of online recovery control for systems with unknown dynamics.
To handle unknown dynamics, we mainly focus on learning-based approaches, since they are independent of system dynamics, and can be adaptable to system changes, which is well-suited for recovery control and can be easily combined with the proposed funnel-based method.
In this paper, a specific type of neural network called Recurrent Equilibrium Network (REN) \cite{revay2023recurrent} is applied since this type of neural network can not only approximate the unknown dynamics online around the nominal trajectories, but also provide input-output properties of the learned unknown dynamics during the training phase.

% 4. Our proposed method

Therefore, we apply the RENs to learn the unknown dynamics online and correct the system dynamics, where we require that the states of the system with the learned unknown dynamics are close to the states of the actual system.
Considering the system with the learned unknown dynamics, we propose a funnel-based recovery control framework so that the closed-loop system is both internally and $L_2$ stable, with invariant funnels obtained along the nominal trajectory. 
% {\color{gray} Even though methods like tube-based MPC \cite{hill2023tube} can ensure the invariance of the system states around the nominal trajectory under unknown dynamics, they may not be tractable for complex nonlinear systems and lack formal guarantees for stability and transition performance during the system recovery. 
% Compared to the above methods, funnel-based control \cite{reynolds2021funnel,kim2023optimization}, as a relatively new method for path planning, appears to be specifically suitable for system recovery control, since multiple constraints such as convergence behaviors, input saturation and safety constraints can be combined in the solutions and the invariant funnels can be systematically derived that allow the maximal state deviations with respect to nominal trajectories.}

% {\color{red} The main contribution of this paper is threefold. First, ... Second, ... Third...}
Based on the above discussions, the main contributions of this paper can be summarized as follows:
\begin{enumerate}
    \item A funnel-based method is proposed for system recovery control with unknown dynamics;
    \item The internal and $L_2$ stability of the closed-loop system under our proposed controller are shown together with the invariant funnel derivation;
    \item Different from the existing work of funnel-based control \cite{reynolds2021funnel,kim2023optimization}, where external disturbances are norm bounded by $1$, our results remove this assumption to handle recovery control scenarios.
\end{enumerate}

% 5. Arrangements
The remainder of this paper is organized as follows. Notations and problem formulation are presented in Section \ref{sec:background}. The REN-based unknown dynamics approximation is presented in Section \ref{sec:REN_unknown_dynamics_approximation}.
Our main results of funnel-based recovery control are presented in Section \ref{sec:funnel-based_online_recovery_control_design}, and are supported by a simulation example in Section \ref{sec:simulation}. Finally, concluding remarks are provided in Section \ref{sec:conclusion}.

%----------------------------------------------------------
% \vspace{-2mm}
\section{Background}\label{sec:background}
%----------------------------------------------------------

\subsection{Notations}

The sets of real, non-negative real, positive real, and natural numbers are denoted by $\mathbb{R}$, $\mathbb{R}_{\geq 0}$, $\mathbb{R}_+$, and $\mathbb{N}$, respectively. 
$\mathbb{N}_0:=\mathbb{N}\cup\{0\}$ is the set of natural numbers containing zero.
$\mathbb{R}^{n\times m}$ denotes the space of real matrices with $n$ rows and $m$ columns. 
$\mathbb{R}_{\geq 0}^{n\times n}$ ($\mathbb{R}_+^{n\times n}$) represents the space of semi-positive (positive definite) matrices with $n$ rows and columns. Similarly, $\mathbb{R}_{\leq 0}^{n\times n}$ represents the space of semi-positive matrices with $n$ rows and columns.
An $n$-dimensional real vector is denoted by $\mathbb{R}^n$. 
% $\mathbb{R}^{n}_+$ denotes the sets that contain all component-wise nonnegative vectors in $\mathbb{R}^n$.
$\mathcal{I}_N:=\{1, 2,...,N\}$ is the index sets, where $N\in \mathbb{N}$, and $\mathcal{I}_N^0:=\mathcal{I}_N\cup\{0\}$.
The zero and identity matrices are denoted by $\0$ and $\I$, respectively (dimensions will be obvious from the context). 
% A block matrix $A\in\mathbb{R}^{n\times m}$ is represented as $A:=[A_{ij}]_{i\in\mathcal{I}_n, j\in\mathcal{I}_m}$, where $A_{ij}$ is the $(i,j)$\tsup{th} block of $A$. $[A_{ij}]_{j\in \mathcal{I}_m}$ represents a block row matrix.
% For block matrix $A=[A_{ij}]_{i,j\in\mathcal{I}_n}$, $A_{[:i,:i]}$ (or $A^{[:i,:i]}$) represents the first $i$ block rows and $i$ block columns extracted from $A$.
% $\mbox{diag}(a)\in\mathbb{R}^{n\times n}$ is a diagonal matrix generated by $a=[a_i]^\T\in\mathbb{R}^n$, where each scalar $a_i$ is on its diagonal (dimensions will be obvious from the context).
% For a vector matrix $X=[x_i]_{i\in\mathcal{I}_N}^\T\in\mathbb{R}^{N\times n}$ with each $x_i\in\mathbb{R}^n$, we define $\mbox{vec}(X^\T):=[x_i^{\top}]_{i\in\mathcal{I}_N}^{\top}\in\mathbb{R}^{nN}$ as its vectorized form.  
% $\mb{1}_{\{\cdot\}}$ is the indicator function and $\mathsf{e}_{ij} \triangleq \mb{1}_{\{i=j\}}$. 
% For two vectors $x$, $y\in\mathbb{R}^n$, we use $x\preceq y$ ($x\succeq y$) to indicate that every element of $x-y$ is non-positive (non-negative), i.e., $x_i\leq y_i$ ($x_i\geq y_i$) for all $i\in \mathcal{I}_n$.
% For matrix $A\in\mathbb{R}^{n\times m}$, its spectral norm is denoted by $\|A\|$.
The set of sequences $x:\mathbb{N}_0\rightarrow\mathbb{R}^n$ is denoted by $\ell_{2e}^n$. For $x\in\ell_{2e}^n$, $x(t)\in\mathbb{R}^n$ is the value of the sequence $x$ at time $t\in\mathbb{N}_0$, and its Euclidean norm is given by $|x(t)|_2 := |x(t)| := \sqrt{x^\T(t) x(t)}$.
Given a sequence $x\in\ell_{2e}^n$, its $L_2$ and $L_{\infty}$ norm over the time interval $[t_0,t_f]$ are $\|x(\cdot)\|:=\sqrt{\sum_{\tau=t_0}^{t_f}|x(\tau)|^2}$ and $\|x(\cdot)\|_{\infty}:=\sup_{\tau\in[t_0,t_f]}\ |x(\tau)|$, respectively. 
The subset $\ell_2^n\subset\ell_{2e}^n$ consists of all square-summable sequences, i.e., $x\in\ell_2^n$ iff $\|x(\cdot)\|:=\sqrt{\sum_{\tau=0}^{\infty} |x(\tau)|^2}<\infty$.
In this paper, we will use $(\cdot)(t)$ or $(\cdot)_t$ interchangeably for time dependent variables according to the context.

% For a vector $x\in\mathbb{R}^n$, its Euclidean norm is given by $|x|_2 := |x| = \sqrt{x^{\top}x}$. 

% For a time dependent sequence $x(t):\mathbb{N}_0\rightarrow\mathbb{R}^n$, $\ell_2^n$ is the set of sequences $x(t)$ with $\|x(t)\|_2:=\sqrt{\sum_{\tau=0}^{\infty} x^\T(\tau)x(\tau)}<\infty$.    

% The $\mathcal{L}_2$ and $\mathcal{L}_{\infty}$ vectored function norms are given by $\|x(\cdot)\|=\sqrt{\int_{0}^{\infty}|x(t)|^2dt}$ and $\|x(\cdot)\|_{\infty} = \sup_{t\geq 0} |x(t)|$, respectively. $\mathcal{K}$ denotes class-$\mathcal{K}$ functions \cite{sontag1995characterizations}.

%------------------------------------------------------
% \vspace{-1mm}
\subsection{Problem Formulation}

Consider the dynamics of a discrete-time nonlinear system as follows:
\begin{equation}\label{Eq:NonlinearSystemDynamics}
        x(t+1) = \underbrace{f(t, x(t), u(t))}_{\mathrm{nomimal\ dyn.}} + \underbrace{\Delta(t)}_{\mathrm{unknown\ dyn.}}\hspace{-5mm},\ \ \forall t\in\mathcal{I}_{N-1}^0,
\end{equation}
where $N\in\mathbb{N}$ is the length of the time steps for the evolution of the system.
% and $t_k$ is the time at the $k\tsup{th}$ step. 
% Here, $(\cdot)_k$ is the signal in the $k\tsup{th}$ step.
$x\in\mathcal{X}\subset\mathbb{R}^n$ and $u\in\mathcal{U}\subset\mathbb{R}^m$ are the system states and inputs, respectively, where $\mathcal{X}:=\{x\in\mathbb{R}^n\ |\ h_i(x)\leq 0,\ i\in\mathcal{I}_{l_x}\}$ and $\mathcal{U}:=\{u\in\mathbb{R}^n\ |\ g_j(u)\leq 0,\ i\in\mathcal{I}_{l_u}\}$ are compact sets for states and inputs, respectively, where all the functions $h_i$ and $g_j$ are state and input constraints, respectively, which are concave and differentiable at least once, for all $i\in\mathcal{I}_{l_x}$ and $j\in\mathcal{I}_{l_u}$. Here, $l_x$, $l_u\in\mathbb{N}$ are numbers of the state and input constraints, respectively.
$\Delta\in\mathbb{R}^n$ is the unknown dynamics caused by attacks or faults.
$f:\mathbb{N}_0\times\mathbb{R}^{n}\times\mathbb{R}^{m}\rightarrow\mathbb{R}^{n}$ is the nonlinear mapping of the system dynamics constructed with prior knowledge of the physics, i.e., the nominal system dynamics, which is continuously differentiable.
It is also assumed that at the time step $t$, for every $x^*\in\mathcal{X}$, there is a unique $u^* \in\mathcal{U}$ that satisfies $f(t, x^*, u^*)=\0$, while $u^*$ is the implicit function of $x^*$. 
In other words, $(x^*,u^*)$ is the equilibrium point of the nominal dynamics $f$ at the time step $t$.
Here, the unknown dynamics $\Delta$ are assumed to be time-varying and bounded. 

% \begin{remark}
%     % Suppose that there exists some prior designed controller $u_p(x)\in\mathcal{U}$ that can stabilize the nominal dynamics $f$ without the unknown dynamics $\Delta$, i.e., the states $x\rightarrow x^*$ as $t\rightarrow\infty$ under the prior designed controller $u_{p}$. 
%     Note that the nonlinear mapping $f$ in \eqref{Eq:NonlinearSystemDynamics} can also represent the dynamics under some prior designed controller $u_p(x)\in\mathcal{U}$, where $u_p$ can stabilize the nominal dynamics $f$ without the unknown dynamics $\Delta$, i.e., $x\rightarrow x^*$ as $t\rightarrow\infty$ under $u_{p}$. 
%     However, due to the attacks or failures, the nominal dynamics $f$ have been changed, and these impacts on $f$ are lumped into the term $\Delta$ as in \eqref{Eq:NonlinearSystemDynamics}.
%     In this case, the prior designed controller $u_{p}$ may no longer stabilize the nominal dynamics $f$.
%     If not properly handled, the performance of the nonlinear system \eqref{Eq:NonlinearSystemDynamics} may be compromised or even unstable. 
%     Now, if we change our actual control input $u(t)$ in \eqref{Eq:NonlinearSystemDynamics} into $u:=u_{p}+\bar{u}$, where $\bar{u}$ is the new control input to be further determined, then the system dynamics under $u_p$ can be absorbed into the nominal dynamics $f$ and the nominal dynamics can be represented by a new mapping $\bar{f}(t,x(t),\bar{u})$, which is of the same form as that in \eqref{Eq:NonlinearSystemDynamics}.
% \end{remark}

Note that the nonlinear mapping $f$ in \eqref{Eq:NonlinearSystemDynamics} can also represent the dynamics under some prior designed controller $u_p(x)\in\mathcal{U}$.
Suppose that there exists some prior designed controller $u_{p}(x)\in\mathcal{U}$ (which is known) such that it can stabilize the nominal dynamics $f$ in \eqref{Eq:NonlinearSystemDynamics}, i.e., the states $x\rightarrow x^*$ as $t\rightarrow N-1$ under the prior designed controller $u_{p}$. 
However, due to the attacks or failures, the nominal dynamics $f$ have been changed, and these impacts on $f$ are lumped into the term $\Delta$ as in \eqref{Eq:NonlinearSystemDynamics}. 
% Here, the unknown dynamics $\Delta$ is assumed to be time-varying and bounded. 
In this case, the prior designed controller $u_{p}$ may no longer stabilize the nonlinear system \eqref{Eq:NonlinearSystemDynamics} under attacks or failures.
If not properly handled, the performance of the nonlinear system \eqref{Eq:NonlinearSystemDynamics} may be compromised or even unstable.

Now, if we change our actual control input $u(t)$ in \eqref{Eq:NonlinearSystemDynamics} to $u:=u_{p}+u_r$, where $u_r\in\mathcal{U}$ is some additional controller to be further determined for system recovery, 
and substitute the state-dependent prior designed controller $u_p$ into the system dynamics \eqref{Eq:NonlinearSystemDynamics}, its dynamics can be rewritten as:
\begin{equation}\label{Eq:NonlinearSystemDynamics_withPriorControl}
        x(t+1) = \underbrace{\bar{f}(t, x(t), u_r(t))}_{\begin{array}{c}
    \scriptstyle \mathrm{nom.\ dyn.\ with} \\ 
    \scriptstyle \mathrm{prior\ controller}
    \end{array}} + \underbrace{\Delta(t)}_{\mathrm{unknown\ dyn.}}\hspace{-5mm},\ \ \forall t\in\mathcal{I}_{N-1}^0,
\end{equation}
where the nominal dynamics $f$ under the prior designed controller $u_p$ is represented by the nonlinear mapping $\bar{f}:\mathbb{N}_0\times\mathbb{R}^{n}\times\mathbb{R}^{m}\rightarrow\mathbb{R}^{n}$ and $\bar{f}$ is also continuously differentiable. 
Therefore, the system dynamics \eqref{Eq:NonlinearSystemDynamics_withPriorControl} is of the same form as that in \eqref{Eq:NonlinearSystemDynamics} with the same state and input compact sets $\mathcal{X}$ and $\mathcal{U}$ with respect to the states $x(t)$ and inputs $u_r(t)$, respectively, as those in \eqref{Eq:NonlinearSystemDynamics}.
% $\mathcal{X}:=\{x\in\mathbb{R}^n\ |\ h_i(x)\leq 0,\ i\in\mathcal{I}_{l_x}\}$ and $\mathcal{U}:=\{u\in\mathbb{R}^n\ |\ g_j(u)\leq 0,\ i\in\mathcal{I}_{l_u}\}$ are compact sets for states and inputs, respectively, where all the functions $h_i$ and $g_j$ are state and input constraints, respectively, which are concave and differentiable at least once, for all $i\in\mathcal{I}_{l_x}$ and $j\in\mathcal{I}_{l_u}$. Here, $l_x$, $l_u\in\mathbb{N}$ are numbers of the state and input constraints, respectively.
Here, $u_r$ is the control input of the new dynamics in \eqref{Eq:NonlinearSystemDynamics_withPriorControl} for the recovery of the system.

\begin{figure}[!h]
    \centering
    \includegraphics[width=3in]{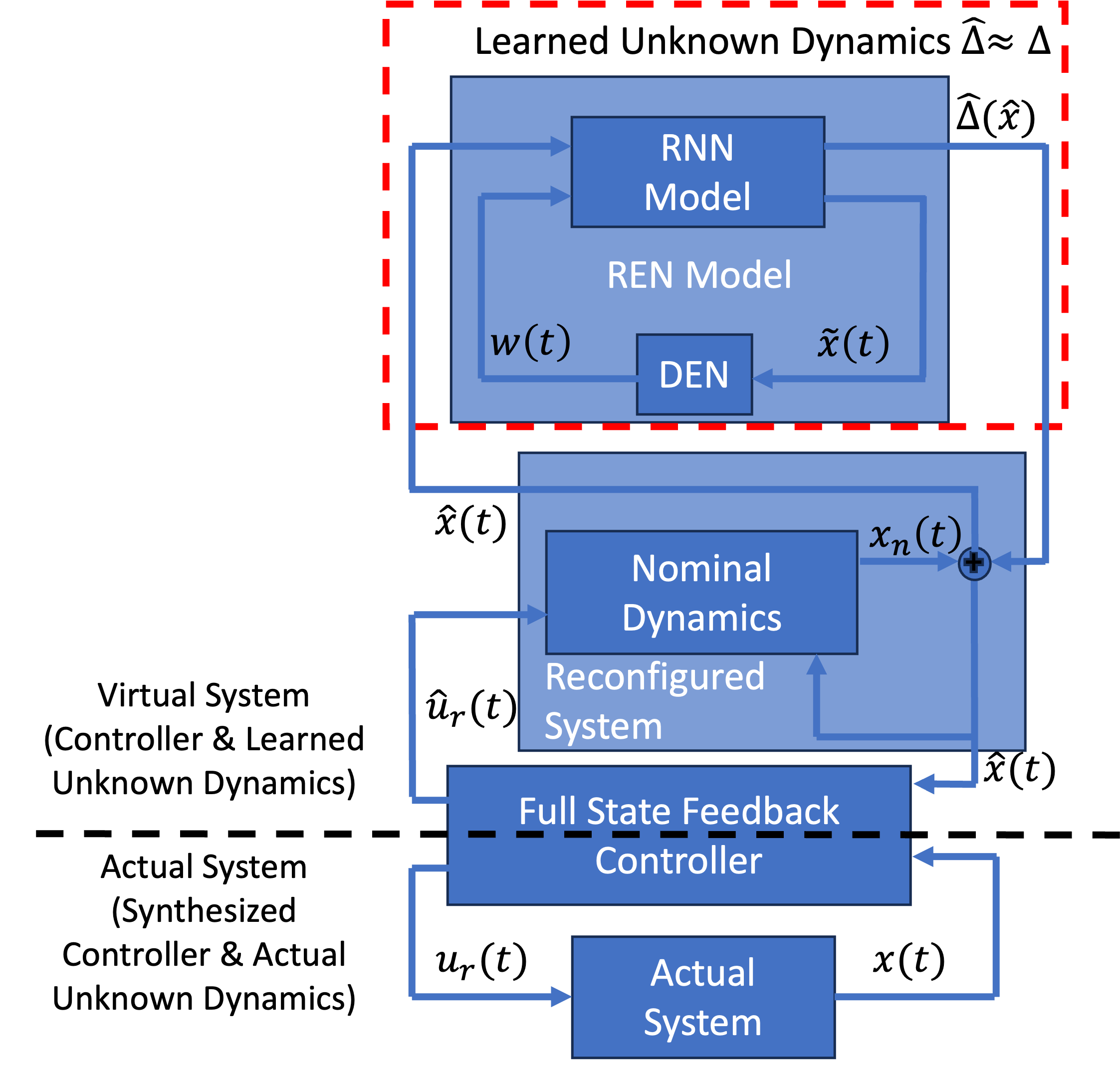}
    \caption{Our proposed recovery control framework, where $x_{n}(t)=A\hat{x}(t)+B\hat{u}_r(t)+Ep(t)$ is the states of the nominal system dynamics. The REN model is trained online using the real-time state data $\hat{x}_{t+1}=x_{nt}+\hat{\Delta}_t$. The actual and the virtual system share the same form of controller, but use different feedback states. }
    \label{Fig:proposed_control_architecture_single_system}
\end{figure}

Based on the above discussion, we aim to design the recovery controller $u_r$ online to return the system states of \eqref{Eq:NonlinearSystemDynamics_withPriorControl} to some neighborhood around the corresponding desired equilibrium points under any time-varying and bounded unknown dynamics $\Delta$. 
% where we apply the Recurrent Equilibrium Networks (RENs) to approximate the . 
% Considering the system dynamics with our learned unknown dynamics, we propose our funnel-based recovery controller, where the internal and $L_2$ stability conditions are shown and the invariant funnels are derived.
The basic architecture of our proposed control framework is shown in Fig. \ref{Fig:proposed_control_architecture_single_system}.

%----------------------------------------------------------------------
\section{REN-based Unknown Dynamics Identification}\label{sec:REN_unknown_dynamics_approximation}

To achieve recovery control of the system \eqref{Eq:NonlinearSystemDynamics_withPriorControl}, we first approximate the time-varying and bounded unknown dynamics $\Delta(t)$ in \eqref{Eq:NonlinearSystemDynamics_withPriorControl} using REN model \cite{revay2023recurrent} to obtain a learned unknown dynamics $\hat{\Delta}(t)\approx \Delta(t)$.
However, the definition of the REN model in \cite{revay2023recurrent} may easily be misinterpreted. Motivated by this reason, and for self-containment of this paper, in this section, we clarify the definition of the REN model.

A REN model is a combination of a Recurrent Neural Network (RNN) \cite{salem2022recurrent} and a Deep Equilibrium Network (DEN) \cite{bai2019deep} (as shown in Fig. \ref{Fig:proposed_control_architecture_single_system}) with formal input-output guarantees, e.g., contraction and incremental $(Q,S,R)$-dissipativity \cite{revay2023recurrent}. Because of the existence of the DEN, the REN model is powerful in terms of highly nonlinear dynamic behavioral learning around the equilibrium points. 
Moreover, fast convergence speed is ensured since RENs are architecturally designed as contractive dynamical systems that directly converge to equilibrium points rather than generic recurrent networks, and thus, enables online implementation. 
Consequently, REN models are suitable for our proposed funnel-based recovery control design (introduced in the next section) that requires linearization of the nonlinear system dynamics \eqref{Eq:NonlinearSystemDynamics_withPriorControl} around the nominal trajectories.

To ensure the input-output properties of the REN model, we first review the concept of incremental integral quadratic constraints (IQCs). In particular, we consider the following (learnable) system:
\begin{align}\label{Eq:general_nonlinear_dynamics}
    \begin{cases}
        x(t+1)=f_{\theta}(x(t),u(t), \theta),    \\
        y(t) = g_{\theta}(x(t),u(t), \theta),
    \end{cases}
\end{align}
where $x\in\mathbb{R}^n$, $u\in\mathbb{R}^m$, and $y\in\mathbb{R}^q$ are system states, inputs, and outputs of the system, respectively;  
$\theta\in\Theta\subseteq\mathbb{R}^p$ are the learnable parameters of the system;  
$f_{\theta}:\mathbb{R}^n\times\mathbb{R}^m\times\Theta\rightarrow\mathbb{R}^n$ and $g_{\theta}:\mathbb{R}^n\times\mathbb{R}^m\times\Theta\rightarrow\mathbb{R}^q$ are piecewise continuously differentiable functions, representing system dynamics and output mappings, respectively.

\begin{definition}(\textit{Incremental IQC \cite{revay2023recurrent}})\label{Def:incremental_IQC}
    % For all pairs of solutions of the system \eqref{Eq:general_nonlinear_dynamics} with initial conditions $a$, $b\in\mathbb{R}^n$ and input sequences $u^a$, $u^b\in\ell_{2e}^m$, the corresponding output sequences $y^a$, $y^b$ generated by the system \eqref{Eq:general_nonlinear_dynamics}
    The system \eqref{Eq:general_nonlinear_dynamics} satisfies the incremental IQC property characterized by $(Q,S,R)$, where $Q=Q^\T\in\mathbb{R}_{\leq 0}^{q\times q}$, $S\in\mathbb{R}^{m\times q}$ and $R=R^\T\in\mathbb{R}^{m\times m}$ from input deviations $u^a-u^b$ to output deviations $y^a-y^b$, if for all pairs of solutions with initial conditions $a$, $b\in\mathbb{R}^n$ and input sequences $u^a$, $u^b\in\ell_{2e}^m$, the corresponding output sequences $y^a$, $y^b$ generated by the system \eqref{Eq:general_nonlinear_dynamics} satisfy
    \begin{align}\label{Eq:incremental_IQC}
        \sum_{t=0}^T\bmtx{y^a(t)-y^b(t) \\u^a(t)-u^b(t)}^\T\bmtx{Q & S^\T \\ S & R}\bmtx{y^a(t)-y^b(t) \\u^a(t)-u^b(t)}\geq -d(a,b),
    \end{align}
    for all $T\in\mathbb{N}_0$ and some function $d(a,b)\geq 0$ with $d(a,a)=0$.
    % where $y^a:=\mathcal{R}_a(u)$ and $y^b:=\mathcal{R}_b(v)$ are output sequences generated by system \eqref{Eq:general_nonlinear_dynamics} with initial conditions $a$ and $b$, respectively, under inputs $u(t)$ and $v(t)$, respectively.
\end{definition}

\begin{remark}
    By different choices of $Q$, $S$ and $R$ matrices, we can ensure different properties of the system \eqref{Eq:general_nonlinear_dynamics}, such as incremental $L_2$-stability, incremental (strict input or output) passivity \cite{revay2023recurrent}. Besides, classical IQC definition \cite{wu2022stability} can be viewed as a special case of the incremental IQC defined in Def. \ref{Def:incremental_IQC}, where $a$ and $y^a$ are selected as zero vectors, the block matrix constructed by $(Q,S,R)$ belongs to some positive semi-definite cone and $d(a, b)\equiv 0$.
\end{remark}

Now, taking the states of the virtual system (as seen in Fig. \ref{Fig:proposed_control_architecture_single_system}) as the input, we construct the REN to approximate the unknown dynamics $\Delta$ in \eqref{Eq:NonlinearSystemDynamics_withPriorControl} as
\begin{subequations}\label{Eq:NodeREN_subsystem_modeling}
    \begin{gather}
        \tilde{x}(t+1) = A\tilde{x}(t)+B_{1}w(t)+B_{2}\hat{x}(t)+b_{x},    \label{Eq:state_transition}    \\
    \hat{\Delta}(t)=C_{2}\tilde{x}(t)+D_{21}w(t)+D_{22}\hat{x}(t)+b_{y},  \label{Eq:learned_output}
    \end{gather}
\end{subequations}
in which $\tilde{x}\in\mathbb{R}^{n_x}$ is the hidden state of REN, $\hat{\Delta}\in\mathbb{R}^n$ is the output of the REN, which is also the approximation of the unknown dynamics $\Delta$ in \eqref{Eq:NonlinearSystemDynamics_withPriorControl}. Here, the virtual state $\hat{x}(t)\in\mathbb{R}^n$ is the input data of the REN model from the virtual system dynamics (shown in Fig. \ref{Fig:proposed_control_architecture_single_system}, which will be introduced later in Section \ref{subsec:closed-loop_dynamics_single}), and $w(t)\in\mathbb{R}^{n_v}$ is the solution of an equilibrium network, also known as the implicit network, which is designed as:
\begin{align}\label{Eq:implicit_network}
    v(t)&=D_{11}w(t)+C_{1}\tilde{x}(t)+D_{12}\hat{x}(t)+b_{v},  \nonumber \\
    w(t) &= \sigma_w(v(t)),
\end{align}
where $A\in\mathbb{R}^{n_x\times n_x}$, $B_1\in\mathbb{R}^{n_x\times n_v}$, $B_2\in\mathbb{R}^{n_x\times n}$, $C_1\in\mathbb{R}^{n_v\times n_x}$, $C_2\in\mathbb{R}^{n\times n_x}$, $D_{11}\in\mathbb{R}^{n_v\times n_v}$, $D_{12}\in\mathbb{R}^{n_v\times n}$, $D_{21}\in\mathbb{R}^{n\times n_v}$ and $D_{22}\in\mathbb{R}^{n\times n}$ are trainable matrix parameters for REN dynamics, $b_{x}\in\mathbb{R}^{n_x}$, $b_{y}\in\mathbb{R}^n$, and $b_{v}\in\mathbb{R}^{n_v}$ are bias vectors, and $\sigma_w$ is an element-wise nonlinear activation function.

Thus, we can rewrite the REN model \eqref{Eq:NodeREN_subsystem_modeling}-\eqref{Eq:implicit_network} in a compact form as:
\begin{subequations}\label{Eq:state_transition_compact}
    \begin{align}
        \bmtx{\tilde{x}(t+1) \\ v(t) \\ \hat{\Delta}(t)} &= \underbrace{\bmtx{A & B_{1} & B_{2} \\ C_{1} & D_{11} & D_{12} \\ C_{2} & D_{21} & D_{22}}}_{\tilde{A}}\bmtx{\tilde{x}(t) \\  w(t) \\ \hat{x}(t)}+\underbrace{\bmtx{b_{x} \\ b_{v} \\ b_{y}}}_{\tilde{b}},    \label{Eq:compact_REN_state_transition}  \\
        w(t) &= \sigma_w(v(t)),  \label{Eq:compact_nonlinear_mapping}
\end{align}
\end{subequations}
where the input and output of this REN model are $\hat{x}$ and $\hat{\Delta}$, respectively. 
The first two equations of $\tilde{x}$ and $v$ in \eqref{Eq:compact_REN_state_transition} and the equation \eqref{Eq:compact_nonlinear_mapping} formulate the dynamic mapping $f_{\theta}$ like that in \eqref{Eq:general_nonlinear_dynamics}, while the equation $\hat{\Delta}$ formulates the output mapping $g_{\theta}$ like that in \eqref{Eq:general_nonlinear_dynamics}, where the trainable parameters in $f_{\theta}$ and $g_{\theta}$ corresponding to the REN model \eqref{Eq:state_transition_compact} are $(A, B_1, B_2, C_1, D_{11}, D_{12}, b_x, b_v)$, and $(C_2, D_{21}, D_{22}, b_y)$, respectively.
% In particular, the REN model takes the output of the $i\tsup{th}$ actual subsystem \eqref{Eq:origin_linearized_subsystem_i} as the input data for the training.

To deal with the nonlinearities elements in \eqref{Eq:state_transition_compact}, we use the IQC and we make the following assumption.
\begin{assumption}\label{ass:slope_restriction_NodeREN}
    The (nonlinear) activation function $\sigma_w$ is piecewise differentiable and slope restricted in $[0,1]$, i.e.,
    \begin{equation}
        0\leq \frac{\sigma_w(y)-\sigma_w(x)}{y-x}\leq 1,\ 
    \end{equation}
    for all $x$, $y\in\mathbb{R},\ x\neq y$.
\end{assumption}
\begin{remark}
    Note that Asm. \ref{ass:slope_restriction_NodeREN} is a typical assumption to limit the output of nonlinear functions. This assumption is not generally restrictive, since it is satisfied for the most frequently used activation functions generally applied in existing work, such as the logistic, ReLu, and tanh activation functions. 
\end{remark}

With Asm. \ref{ass:slope_restriction_NodeREN}, we have:
\begin{equation}\label{Eq:slope_restriction_sigma_w}
    \Gamma(t):=\bmtx{\Delta v \\ \Delta w}^\T\bmtx{\0 & \Lambda_{w} \\ \Lambda_{w} & -2\Lambda_{w}}\bmtx{\Delta v \\ \Delta w} \geq 0, 
\end{equation}
for any positive diagonal matrix $\Lambda_w>0$.

To ensure incremental IQC property defined by ($Q,S,R$) from input deviations $\hat{x}^a-\hat{x}^b$ to output deviations $\hat{\Delta}^a-\hat{\Delta}^b$ of the REN model, we present the following theorem.
\begin{theorem}\cite{revay2023recurrent}\label{th:incremental_IQC_REN}
    The REN model \eqref{Eq:state_transition_compact} satisfies the incremental IQC from input deviations $\hat{x}^a-\hat{x}^b$ to output deviations $\hat{\Delta}^a-\hat{\Delta}^b$  characterized by ($Q,S,R$) and is well-posed, if for a given $\bar{\alpha}\in(0,1]$, there exists $P_{1}>0$ and a diagonal matrix $\Lambda_{w}>0$, such that
    \begin{align}\label{Eq:incremental_IQC_REN}
        & \bmtx{\bar{\alpha}^2 P_1 & -C_1^\T\Lambda_w & C_2^\T S^\T \\ -\Lambda_w C_1 & F & D_{21}^\T S^\T-\Lambda_w D_{12} \\ 
        SC_2 & SD_{21}-D_{12}^\T\Lambda_w & R+S D_{22}+D_{22}^\T S^\T}- \nonumber \\
        & \bmtx{A^\T \\ B_1^\T \\ B_2^\T} P_1\bmtx{A^\T \\ B_1^\T \\ B_2^\T}^\T+\bmtx{C_{2}^\T \\ D_{21}^\T \\ D_{22}^\T}Q\bmtx{C_{2}^\T \\ D_{21}^\T \\ D_{22}^\T}^\T > 0
    \end{align}
    holds, where $F:=2\Lambda_{w}-\Lambda_{w}D_{11}-D_{11}^\T\Lambda_{w}>0$ is the well-posedness condition of the REN model in \eqref{Eq:state_transition_compact}.
    % \begin{equation}\label{Eq:incremental_IQC_REN_F}
    %     F:=2\Lambda_{w}-\Lambda_{w}D_{11}-D_{11}^\T\Lambda_{w}.
    % \end{equation}
\end{theorem}

The condition \eqref{Eq:incremental_IQC_REN} is embedded in the training phase of the unknown dynamics $\Delta$ modeled by REN (see Fig. \ref{Fig:proposed_control_architecture_single_system}). Specific details on the convexification of \eqref{Eq:incremental_IQC_REN} and its direct parameterization can be found in \cite{revay2023recurrent}.

\begin{remark}
    For several frequently used incremental properties \cite[Def. 3]{revay2023recurrent}, we usually have $Q=Q^\T\leq 0$ and $R=R^\T$.
    With these choices, the well-posedness condition $F>0$ can also be enforced through \eqref{Eq:incremental_IQC_REN}.
\end{remark}
\begin{remark}
    The condition \eqref{Eq:incremental_IQC_REN}, particularly the corresponding IQC property, will be exploited in the sequel for our approximation of the unknown dynamics $\Delta(t)$. This is because our proposed control architecture (as seen in Fig. \ref{Fig:proposed_control_architecture_single_system}) follows a similar structure as the classical robust control (e.g., $H_{\infty}$ control \cite{glover2005state}), where the system uncertainties (or unknown dynamics) are required to be stable with certain $H_{\infty}$ norm in frequency domain (also the $L_2$-gain in time domain). This is exactly what Thm. \ref{th:incremental_IQC_REN} formally guarantees if we select $Q=-\frac{1}{\gamma_2}\I$, $S=\0$ and $R=\gamma_2\I$, where $\gamma_2\in\mathbb{R}_+$ is the $L_2$ gain from the input $\hat{x}$ to the output $\hat{\Delta}$. 
    Compared to the classical $H_{\infty}$ control, the unknown dynamics $\hat{\Delta}(t)$ learned by RENs are less conservative, since we do not require the unknown dynamics entering the system in linear fractional transformation (LFT) form or being time-invariant or slowly time-varying. Instead, the $L_2$-gain of the REN-based model is tunable, and thus, more flexible.
\end{remark}

With the learned unknown dynamics $\hat{\Delta}$ by REN model \eqref{Eq:state_transition_compact}, we can reconfigure a virtual system dynamics with the same nonlinear dynamic mapping $\bar{f}$ as the actual dynamics \eqref{Eq:NonlinearSystemDynamics_withPriorControl}, but with different system states and inputs as follows:
\begin{align}\label{Eq:virtual_system_learned_unknown_dynamics}
    \hat{x}(t+1) = \bar{f}(t, \hat{x}(t), \hat{u}_r(t)) + \hat{\Delta}(t),\ \ \forall t\in\mathcal{I}_{N-1}^0,
\end{align}
where $\hat{x}(t)\in\hat{\mathcal{X}}\subset\mathbb{R}^n$ and $\hat{u}_r(t)\in\hat{\mathcal{U}}\subset\mathbb{R}^m$ are the states and control inputs of this virtual system, respectively.
Similar to the setups of the actual system \eqref{Eq:NonlinearSystemDynamics_withPriorControl}, the compact state and input spaces are $\hat{\mathcal{X}}:=\{\hat{x}\in\mathbb{R}^n\ |\ h_i(\hat{x})\leq 0,\ i\in\mathcal{I}_{l_x}\}$ and $\hat{\mathcal{U}}:=\{\hat{u}_r\in\mathbb{R}^m\ |\ g_j(\hat{u}_r)\leq 0,\ j\in\mathcal{I}_{l_u}\}$, respectively, with all the functions $h_i$ and $g_j$ the same as the actual system \eqref{Eq:NonlinearSystemDynamics_withPriorControl}. 
% , for all $i\in\mathcal{I}_{l_x}$ and $j\in\mathcal{I}_{l_u}$. 
% $l_x, l_u\in\mathbb{N}$ are numbers of surfaces to construct these two feasible regions, respectively.

The REN model is trained using the real-time states of the virtual system, and thus, the training of REN is self-supervised and the loss function is selected by minimizing the difference between the actual system states $x(t)$ and the virtual system states $\hat{x}(t)$, i.e.,
\begin{align}
    L_{\mathrm{REN}}:=\frac{1}{nN_{\mathrm{batch}}}\sum_{k=1}^{N_{\mathrm{batch}}}\sum_{i=1}^{n}|x_{ki}(t)-\hat{x}_{ki}(t)|^2,
\end{align}
where $x_{ki}(t)$ is the $i\tsup{th}$ component of the real-time state $x(t)$ of the actual system dynamics \eqref{Eq:NonlinearSystemDynamics_withPriorControl} in the $k\tsup{th}$ batch, and similarly, $\hat{x}_{ki}(t)$ is the $i\tsup{th}$ component of the states $\hat{x}(t)$ of the virtual system 
% governed by the dynamics $\hat{x}(t+1)=\bar{f}(t,\hat{x}(t),\hat{u}_r(t))+\hat{\Delta}(t)$, for all $t\in\mathcal{I}_{N-1}^0$, which is with the same dynamic mapping $\bar{f}$ as that in \eqref{Eq:NonlinearSystemDynamics_withPriorControl} but with the learned unknown dynamics $\hat{\Delta}$, 
in the $k\tsup{th}$ batch.

%----------------------------------------------------------------------
\section{Funnel-based Online Recovery Control Design}\label{sec:funnel-based_online_recovery_control_design}

%----------------------------------------------------------------------
In this section, we propose our main theoretical results of our funnel-based recovery control design for the system with learned unknown dynamics and derive the invariant funnels for the states of the system under our designed controller.

%----------------------------------------------------------------------
\subsection{Closed-Loop Dynamics}\label{subsec:closed-loop_dynamics_single}

With the learned unknown dynamics $\hat{\Delta}$ using the REN model in \eqref{Eq:state_transition_compact}, we can linearize the virtual nonlinear system \eqref{Eq:virtual_system_learned_unknown_dynamics} around this nominal trajectory and convert the virtual system \eqref{Eq:virtual_system_learned_unknown_dynamics} into the following Lur'e type dynamics \cite{boyd1994linear}:
\begin{align}\label{Eq:Lur'e_type_system}
    \hspace{-2mm}\begin{cases}
        \hat{x}(t+1)= A(t)\hat{x}(t)+B(t)\hat{u}_r(t)+ Ep(t)+\hat{\Delta}(t),    \\
        p(t)=\phi(t, q(t)),    \\
        q(t)=C\hat{x}(t)+D\hat{u}_r(t)+G\hat{\Delta}(t),
    \end{cases}
\end{align}
where $\hat{x}\in\mathbb{R}^n$ and $\hat{u}_r\in\mathbb{R}^m$ are the states and control input of the Lur'e dynamics with the learned unknown dynamics $\hat{\Delta}\in\mathbb{R}^n$, which are the same as those in \eqref{Eq:virtual_system_learned_unknown_dynamics}. 
$p\in\mathbb{R}^{n_p}$ is the nonlinear term to compensate the linearization errors, which is a time-varying function with argument $q\in\mathbb{R}^{n_q}$.
This nonlinear function $\phi:\mathbb{N}_0\times\mathbb{R}^{n_q}\rightarrow\mathbb{R}^{n_p}$, which is continuously differentiable, captures the nonlinearities and the dominant linearization errors of the virtual system \eqref{Eq:virtual_system_learned_unknown_dynamics}.
$A(t)=\frac{\partial \bar{f}}{\partial \hat{x}}\big|_{(\bar{x}, \bar{u}_r)}\in\mathbb{R}^{n\times n}$ and $B(t)=\frac{\partial \bar{f}}{\partial \hat{u}_r}\big|_{(\bar{x}, \bar{u}_r)}\in\mathbb{R}^{n\times m}$ are the first-order approximations of the nonlinear system \eqref{Eq:NonlinearSystemDynamics_withPriorControl} around the nominal trajectory. 
The matrices $E\in\mathbb{R}^{n\times n_p}$, $C\in\mathbb{R}^{n_q\times n}$, $D\in\mathbb{R}^{n_q\times m}$ and $G\in\mathbb{R}^{n_q\times n}$ are time-invariant selector matrices with $0$'s and $1$'s to organize the nonlinearity of the system.

\begin{remark}
    Note that the learned unknown dynamics $\hat{\Delta}$ in \eqref{Eq:Lur'e_type_system} can not only approximate the actual unknown dynamics caused by attacks or failures, but also compensate for small model mismatches, while the dominant linearization errors are directly modeled by the nonlinear term $p(t)$ in \eqref{Eq:Lur'e_type_system}.
\end{remark}

To design the funnel-based recovery controller for the virtual system \eqref{Eq:virtual_system_learned_unknown_dynamics}, suppose that a nominal trajectory has been obtained. In particular, the dynamics of the nominal trajectory are as follows:
\begin{align}
    \begin{cases}
        \bar{x}(t+1)= A(t)\bar{x}(t)+B(t)\bar{u}_r(t)+ E\bar{p}(t),    \\
        \bar{p}(t)=\phi(t, \bar{q}(t)),    \\
        \bar{q}(t)=C\bar{x}(t)+D\bar{u}_r(t),
    \end{cases}
\end{align}
where $(\bar{x}(\cdot), \bar{u}_r(\cdot))$ defines the nominal trajectory, which is a solution of the virtual system \eqref{Eq:virtual_system_learned_unknown_dynamics} without learned unknown dynamics, i.e., $\hat{\Delta}\equiv \0$. 
More details of designing this trajectory can be found in \cite{kim2024joint}.

Define state and control input tracking errors with respect to the nominal trajectory $(\bar{x}(\cdot), \bar{u}_r(\cdot))$ are $\eta(t):=\hat{x}(t)-\bar{x}(t)$ and $\xi(t):=\hat{u}_r(t)-\bar{u}_r(t)$, respectively. 
Thus, the tracking error dynamics of \eqref{Eq:Lur'e_type_system} with respect to the nominal trajectory $(\bar{x}(\cdot), \bar{u}_r(\cdot))$ can be obtained by
\begin{align}\label{Eq:tracking_error_Lur'e_dynamics}
    \begin{cases}
        \eta(t+1)=A(t)\eta(t)+B(t)\xi(t)+E\tilde{p}(t)+\hat{\Delta}(t),     \\
        \tilde{p}(t)=\phi(t,q(t))-\phi(t,\bar{q}(t)),    \\
        \tilde{q}(t)=C\eta(t)+D\xi(t)+G\hat{\Delta}(t),
    \end{cases}
\end{align}
where $\tilde{p}(t):=p(t)-\bar{p}(t)$ with $\bar{q}(t):=\bar{C}\bar{x}(t)+\bar{D}\bar{u}_r(t)$. 

Since the function $\phi$ is continuously differentiable, they are locally Lipschitz, and the following condition
\begin{align}\label{Eq:local_Lipschitz_condition_single_dynamics}
    |p_t-\bar{p}_t|\leq\gamma_{1t}|q_t-\bar{q}_t|,\ \ \forall q_t,\ \bar{q}_t\in\mathcal{Q},
\end{align}
is satisfied for all $t\in\mathcal{I}_{N-1}^0$, where $\gamma_{1t}\in\mathbb{R}_{+}$ is a Lipschitz constant at each time instant and $\mathcal{Q}\subseteq\mathbb{R}^{n_q}$ is any compact set. The estimation of $\gamma_{1t}$ for all $t\in\mathcal{I}_{N-1}^0$ can be found in \cite{kim2024joint}, which can be solved in real-time.

For tracking error dynamics \eqref{Eq:tracking_error_Lur'e_dynamics}, we design the full-state feedback controller as
\begin{align}\label{Eq:full_state_feedback_controller}
    & \xi(t)=K(t)\eta(t),   \nonumber \\
    \Leftrightarrow\ \ & \hat{u}_r(t)=\bar{u}_r(t)+K(t)(\hat{x}(t)-\bar{x}(t)), 
\end{align}
where $K(t)\in\mathbb{R}^{m\times n}$ is the controller gain to be determined.

\begin{remark}
    When applied to the actual system \eqref{Eq:NonlinearSystemDynamics_withPriorControl}, the same designed controller gains $K(t)$ in \eqref{Eq:full_state_feedback_controller} can be used, i.e., the controller applied in actual system is $u_r(t)=\bar{u}_r(t)+K(t)(x(t)-\bar{x}(t))$. This can also be seen in our proposed control architecture in Fig. \ref{Fig:proposed_control_architecture_single_system}, where both sides share the same form of controller.
\end{remark}

Under the controller \eqref{Eq:full_state_feedback_controller}, the closed-loop tracking error dynamics \eqref{Eq:tracking_error_Lur'e_dynamics} can be derived as:
\begin{align}\label{Eq:tracking_error_closed-loop_Lur'e_dynamics}
    \begin{cases}
        \eta_{t+1}=(A_t+B_tK_t)\eta_t+E\tilde{p}_t+\hat{\Delta}_t,     \\
        \tilde{q}_t=(C+DK_t)\eta_t+G\hat{\Delta}_t,
    \end{cases}
\end{align}
with $|\tilde{p}_t|\leq\gamma_{1t}|\tilde{q}_t|$ (from \eqref{Eq:local_Lipschitz_condition_single_dynamics}). Besides, in the sequel, we determine the controller gain $K(t)$ in \eqref{Eq:full_state_feedback_controller} so that the closed-loop system \eqref{Eq:tracking_error_closed-loop_Lur'e_dynamics} is both internally and $L_2$ stable from the learned unknown dynamics $\hat{\Delta}_t$ to the tracking error state $\eta_t$.

\begin{remark}
    With \eqref{Eq:tracking_error_closed-loop_Lur'e_dynamics}, we represent the closed-loop nonlinear system with the state and input dependent nonlinear term $\tilde{p}$ given in \eqref{Eq:tracking_error_Lur'e_dynamics} as a linear form. 
    This may be a conservative way to handle the nonlinear system, but it allows us to design a quadratic Lyapunov function with which we can guarantee the invariance and attractive conditions required in the funnel-based recovery control method we propose for the actual nonlinear system \eqref{Eq:NonlinearSystemDynamics_withPriorControl}. Besides, the construction of \eqref{Eq:tracking_error_closed-loop_Lur'e_dynamics} enables the analysis using linear approaches, and thus, can be solved efficiently.
    These advantages are important for recovery control problem considered in this paper, since to achieve system recovery, we not only need to design the feedback controller \eqref{Eq:full_state_feedback_controller}, but also require to provide a formal guarantee of the largest deviation domain for the real-time states. 
    This also motivates the proposed funnel-based recovery control method shown next.
\end{remark}

%----------------------------------------------------------------------
\subsection{Stability Analysis of the Closed-Loop System}

% Now, we can design the trajectory stabilizing controller (i.e., the gain $K_t$ in \eqref{Eq:full_state_feedback_controller}) and the corresponding invariant funnels along the nominal trajectory. 

Now, using the trajectory stabilizing controller (i.e., the gain $K_t$ in \eqref{Eq:full_state_feedback_controller}), we can analyze the stability of the closed-loop system \eqref{Eq:tracking_error_closed-loop_Lur'e_dynamics}. 
Different from the existing work \cite{reynolds2021funnel,kim2023optimization}, which assumes that the unknown dynamics satisfy $\|\hat{\Delta}(\cdot)\|_{\infty}\leq 1$, we remove this assumption in our funnel-based recovery control design as shown in the following theorem.
\begin{theorem}\label{Th:L2_stability_main_condition}
    The closed-loop system \eqref{Eq:tracking_error_closed-loop_Lur'e_dynamics} under the controller \eqref{Eq:full_state_feedback_controller} is both internally and $L_2$ stable with the $L_2$-gain $\gamma_2\in\mathbb{R}_+$, 
    % for the given local Lipschitz constant $\gamma_{1t}\in\mathbb{R}_{+}$, 
    if there exists $P_t\in\mathbb{R}_{+}^{n\times n}$, $Y_t\in\mathbb{R}^{m\times n}$, $\alpha\in(0,1]$ and $\nu_t\in\mathbb{R}_+$ such that the following difference matrix inequality (DMI) is satisfied
    \begin{align}\label{Eq:DMI_main_condition}
    \bmtx{\alpha P_t & \0 & \0 & (H_t^1)^\T & (H_t^2)^\T & P_t    \\ 
    (\star)^\T & \nu_t\I & \0 & \nu_t E^\T & \0 & \0     \\ 
    (\star)^\T & (\star)^\T & \gamma_{2}\I & \I & G^\T & \0 \\
    (\star)^\T & (\star)^\T & (\star)^\T & P_{t+1} & \0 & \0    \\ 
    (\star)^\T & (\star)^\T & (\star)^\T & (\star)^\T & \nu_t\frac{1}{\gamma_{1t}^2}\I & \0   \\
    (\star)^\T & (\star)^\T & (\star)^\T & (\star)^\T & (\star)^\T & \gamma_{2}\I}\geq 0,
\end{align}
for all $t\in\mathcal{I}_{N-1}^0$, $H_t^1=A_tP_t+B_tY_t$ and $H_t^2=CP_t+DY_t$, with 
% then the closed-loop system \eqref{Eq:tracking_error_closed-loop_Lur'e_dynamics} is both internally and $L_2$-stable with 
$K_t=Y_tP_t^{-1}$ for the controller in \eqref{Eq:full_state_feedback_controller}. 
\end{theorem}

\begin{proof}
Select the Lyapunov function candidate as follows:
\begin{align}\label{Eq:time-varying_Lyapunov_function}
    V(t,\eta):=\eta^\T(t)P^{-1}(t)\eta(t),
\end{align}
where $P(t)\in\mathbb{R}_{+}^{n\times n}$, for all $t\in\mathcal{I}_{N}^0$. 

Then, if we take the difference between $V_{t+1}$ and $\alpha V_t$ $(\alpha\in(0,1])$ along the closed-loop dynamics \eqref{Eq:tracking_error_closed-loop_Lur'e_dynamics}, we have:
\begin{align}\label{Eq:V_dot1}
    & \Delta V_t =\eta_{t+1}^\T P_{t+1}^{-1}\eta_{t+1}-\alpha\eta_t^\T P_t^{-1}\eta_t  \nonumber   \\
    &=((A_t+B_tK_t)\eta_t+E\tilde{p}_t+\hat{\Delta}_t)^\T P_{t+1}^{-1}((A_t+B_tK_t)\eta_t+         \nonumber   \\
    & E\tilde{p}_t+\hat{\Delta}_t)-\alpha\eta_t^\T P_t^{-1}\eta_t    \nonumber      \\
    % =&\eta^\T((A_t+B_tK_t)^\T P_{t+1}^{-1}(A_t+B_tK_t)+P_t^{-1}(A_t+B_tK_t))\eta+  \nonumber   \\
    % &\tilde{p}^\T E^\T P_t^{-1}\eta+\hat{\Delta}^\T P_t^{-1}\eta+\eta^\T P_t^{-1}E\tilde{p}+\eta^\T P_t^{-1}\hat{\Delta}-   \nonumber  \\
    % &\eta^\T P_t^{-1}\dot{P}_tP_t^{-1}\eta \nonumber \\
    &=\bmtx{\eta_t \\ \tilde{p}_t \\ \hat{\Delta}_t}^\T\bmtx{(A_t^{cl})^\T \\ E^\T \\ \I} P_{t+1}^{-1}\bmtx{(A_t^{cl})^\T \\ E^\T \\ \I}^\T \bmtx{\star}-\alpha\eta_t^\T P_t^{-1}\eta_t,  
    % &=\bmtx{\eta \\ \tilde{p} \\ \hat{\Delta}}^\T\bmtx{\bar{M}_t-P_t^{-1} & (A_t^{cl})^\T P_{t+1}^{-1}E & (A_t^{cl})^\T P_{t+1}^{-1}   \\
    % (\star)^\T & E^\T P_{t+1}^{-1}E & E^\T P_{t+1}^{-1} \\ 
    % (\star)^\T & (\star)^\T & P_{t+1}^{-1}}\bmtx{\star}
\end{align}
for all $t\in\mathcal{I}_{N-1}^0$, where $A_t^{cl}=A_t+B_tK_t$.
% $\bar{M}_t=(A_t^{cl})^\T P_{t+1}^{-1}A_t^{cl}$ with

% Since the nonlinear function $\phi$ is  locally Lipschitz with the Lipschitz constant $\gamma_{1t}$ (as seen in \eqref{Eq:tracking_error_closed-loop_Lur'e_dynamics}), the vectors $\tilde{p}_t$ and $\tilde{q}_t$ are related by the condition 

From \eqref{Eq:tracking_error_closed-loop_Lur'e_dynamics} (particularly, due to \eqref{Eq:local_Lipschitz_condition_single_dynamics}), we have $|\tilde{p}_t|\leq\gamma_{1t}|\tilde{q}_t|$, which can be restated as
$$\bmtx{\eta_t \\ \tilde{p}_t \\ \hat{\Delta}_t}^\T\bmtx{C_t^{cl} & \0 & G \\ \0 & \I & \0}^\T\bmtx{\gamma_{1t}^2\I & \0 \\ \0 & -\I}\bmtx{\star}.$$
Besides, we require the closed-loop system \eqref{Eq:tracking_error_closed-loop_Lur'e_dynamics} to be $L_2$-stable from the learned unknown dynamics $\hat{\Delta}_t$ to the tracking error state $\eta_t$ with the $L_2$-gain $\gamma_2$, i.e., $|\eta_t|\leq\gamma_{2}|\hat{\Delta}_t|$ with $\gamma_{2}\in\mathbb{R}_+$ for all $t\in\mathcal{I}_{N}^0$. 
Under the prior constraint, to enforce the latter requirement, we use the IQCs (as shown in Def. \ref{Def:incremental_IQC}) and S-procedure \cite{polik2007survey}, and for \eqref{Eq:V_dot1}, we require:
\begin{align}\label{Eq:V_dot2}
    & \Delta V_t+\bmtx{\eta_t \\ \tilde{p}_t \\ \hat{\Delta}_t}^\T\bigg(\lambda_{pt}\bmtx{C_t^{cl} & \0 & G \\ \0 & \I & \0}^\T\bmtx{\gamma_{1t}^2\I & \0 \\ \0 & -\I}\bmtx{\star}+  \nonumber \\
    & \bmtx{\gamma_{2}^{-1}\I & \0 & \0 \\ (\star)^\T & \0 & \0 \\ (\star)^\T & (\star)^\T & -\gamma_{2}\I}\bigg)\bmtx{\star}\leq 0, 
\end{align}
for all $t\in\mathcal{I}_{N-1}^0$, where $C_t^{cl}=C+DK_t$, and $\lambda_{pt}\in\mathbb{R}_{+}$.

Substituting \eqref{Eq:V_dot1} into \eqref{Eq:V_dot2}, we equivalently have:
\begin{align}\label{Eq:V_dot2_substitution}
    & \bmtx{(A_t^{cl})^\T \\ E^\T \\ \I} P_{t+1}^{-1}\bmtx{(A_t^{cl})^\T \\ E^\T \\ \I}^\T-\bmtx{\alpha P_t^{-1} & \0 & \0 \\ (\star)^\T & \0 & \0 \\ (\star)^\T & (\star)^\T & \0}+   \nonumber \\
    & \lambda_{pt}\bmtx{C_t^{cl} & \0 & G \\ \0 & \I & \0}^\T\bmtx{\gamma_{1t}^2\I & \0 \\ \0 & -\I}\bmtx{\star}+ \nonumber \\ 
    & \bmtx{\gamma_{2}^{-1}\I & \0 & \0 \\ (\star)^\T & \0 & \0 \\ (\star)^\T & (\star)^\T & -\gamma_{2}\I}\leq 0.
\end{align}

Applying Schur complement for the DMI in \eqref{Eq:V_dot2_substitution}, we have:
\begin{align}
    & \bmtx{\alpha P_t^{-1}-\gamma_{2}^{-1}\I & \0 & \0 & (A_t^{cl})^\T    \\
    (\star)^\T & \lambda_{pt}\I & \0 & E^\T   \\
    (\star)^\T & (\star)^\T & \gamma_{2}\I & \I \\
    (\star)^\T & (\star)^\T & (\star)^\T & P_{t+1}}-     \nonumber    \\ 
    & \lambda_{pt}\gamma_{1t}^2\bmtx{C_t^{cl} & \0 & G & \0}^\T\bmtx{\star}\geq 0,   \nonumber    
\end{align}
\begin{align}\label{Eq:DMI_Schur_complement}
    \Leftrightarrow\ & \bmtx{\alpha P_t^{-1}-\gamma_{2}^{-1}\I & \0 & \0 & (A_t^{cl})^\T  &  (C_t^{cl})^\T   \\
    (\star)^\T & \lambda_{pt}\I & \0 & E^\T & \0  \\
    (\star)^\T & (\star)^\T & \gamma_{2}\I & \I & G^\T \\
    (\star)^\T & (\star)^\T & (\star)^\T & P_{t+1} & \0 \\
    (\star)^\T & (\star)^\T & (\star)^\T & (\star)^\T & \nu_t\frac{1}{\gamma_{1t}^2}\I}\geq 0,   \nonumber   \\
    \Leftrightarrow\ & \bmtx{\alpha P_t^{-1} & \0 & \0 & (A_t^{cl})^\T  &  (C_t^{cl})^\T  & \I    \\
    (\star)^\T & \lambda_{pt}\I & \0 & E^\T & \0  & \0 \\
    (\star)^\T & (\star)^\T & \gamma_{2}\I & \I & G^\T & \0 \\
    (\star)^\T & (\star)^\T & (\star)^\T & P_{t+1} & \0 & \0 \\
    (\star)^\T & (\star)^\T & (\star)^\T & (\star)^\T & \nu_t\frac{1}{\gamma_{1t}^2}\I & \0 \\
    (\star)^\T & (\star)^\T & (\star)^\T & (\star)^\T & (\star)^\T & \gamma_{2}\I}\geq 0,   
\end{align}
for all $t\in\mathcal{I}_{N-1}^0$, where $\nu_t=\lambda_{pt}^{-1}$.

Then, by left and right multiply a diagonal matrix $\diag([P_t,\nu_t\I,\I,\I,\I,\I])$, we can equivalently transform the DMI in \eqref{Eq:DMI_Schur_complement} as:
\begin{align*}
    \bmtx{\alpha P_t & \0 & \0 & P_t(A_t^{cl})^\T & P_t(C_t^{cl})^\T & P_t    \\ 
    (\star)^\T & \nu_t\I & \0 & \nu_t E^\T & \0 & \0     \\ 
    (\star)^\T & (\star)^\T & \gamma_{2}\I & \I & G^\T & \0 \\
    (\star)^\T & (\star)^\T & (\star)^\T & P_{t+1} & \0 & \0    \\ 
    (\star)^\T & (\star)^\T & (\star)^\T & (\star)^\T & \nu_t\frac{1}{\gamma_{1t}^2}\I & \0   \\
    (\star)^\T & (\star)^\T & (\star)^\T & (\star)^\T & (\star)^\T & \gamma_{2}\I}\geq 0,
\end{align*}
for all $t\in\mathcal{I}_{N-1}^0$.

Now, if we define $K_t:=Y_t P_t^{-1}$, the above DMI 
% in \eqref{Eq:DMI_Schur_complement_transform} 
can be rewritten as
\begin{align*}
\bmtx{\alpha P_t & \0 & \0 & (H_t^1)^\T & (H_t^2)^\T & P_t    \\ 
    (\star)^\T & \nu_t\I & \0 & \nu_t E^\T & \0 & \0     \\ 
    (\star)^\T & (\star)^\T & \gamma_{2}\I & \I & G^\T & \0 \\
    (\star)^\T & (\star)^\T & (\star)^\T & P_{t+1} & \0 & \0    \\ 
    (\star)^\T & (\star)^\T & (\star)^\T & (\star)^\T & \nu_t\frac{1}{\gamma_{1t}^2}\I & \0   \\
    (\star)^\T & (\star)^\T & (\star)^\T & (\star)^\T & (\star)^\T & \gamma_{2}\I}\geq 0.
\end{align*}
% \begin{align*}
%     \bmtx{M_t-\dot{P}_t & \nu E & \I & P_t C^\T+Y_t^\T D^\T & P_t \\ (\star)^\T & -\nu\I & \0 & \0 & \0 \\ 
%     (\star)^\T & (\star)^\T & -\gamma_{2t}\I & G^\T \\
%     (\star)^\T & (\star)^\T & (\star)^\T & -\nu\frac{1}{\gamma_{1t}^2}\I & \0 \\ 
%     P_t & \0 & \0 & \0 & -\gamma_{2t}\I}\leq 0,
% \end{align*}
for all $t\in\mathcal{I}_{N-1}^0$, where $H_t^1:=A_tP_t+B_tY_t$ and $H_t^2:=CP_t+DY_t$, which is exactly the condition \eqref{Eq:DMI_main_condition}.

Eventually, it is worth noting that the above analysis also shows the internal stability of the closed-loop system \eqref{Eq:tracking_error_closed-loop_Lur'e_dynamics} (i.e., the case when $\hat{\Delta}_t\equiv \0$, for all $t\in\mathcal{I}_{N-1}^0$). 
Note that the condition \eqref{Eq:DMI_main_condition} is equivalent to the condition \eqref{Eq:V_dot2}, which can be rewritten as (when $\hat{\Delta}_t\equiv \0$, for all $t\in\mathcal{I}_{N-1}^0$):
\begin{align}\label{Eq:V_dot2_internal_stability}
    & \Delta V_t +\lambda_{pt}(\gamma_{1t}^2|\tilde{q}_t|^2-|\tilde{p}_t|^2) \leq -\frac{1}{\gamma_{2}}|\eta_t|^2,     \nonumber \\
    \Rightarrow\ &  \Delta V_t\leq 0,\ 
    \Rightarrow\ V_{t+1}\leq \alpha V_t\ \ (\alpha\in(0,1]),
\end{align}
which means that $\eta_t\rightarrow\0$ as $t\rightarrow\infty$, and for finite-time horizon, $\eta_t$ is bounded as $t\rightarrow N$, i.e., the closed-loop system \eqref{Eq:tracking_error_closed-loop_Lur'e_dynamics} is internally stable.
Thus, this completes the proof.
\end{proof}

\begin{remark}
    In real-time control, we need to solve the controller $K(t)$ at each time step using DMI \eqref{Eq:DMI_main_condition}, where we treat each $P_t$, for all $t\in\mathcal{I}_{N-1}^0$, as independent variables. 
    For online designs, we can solve this DMI for one step or a certain horizon of time to obtain the controller $K(t)$ at the current time iteratively.
\end{remark}

%----------------------------------------------------------------------
\subsection{Invariant Funnel Synthesis}

\begin{figure}[!t]
    \centering
    \includegraphics[width=\linewidth]{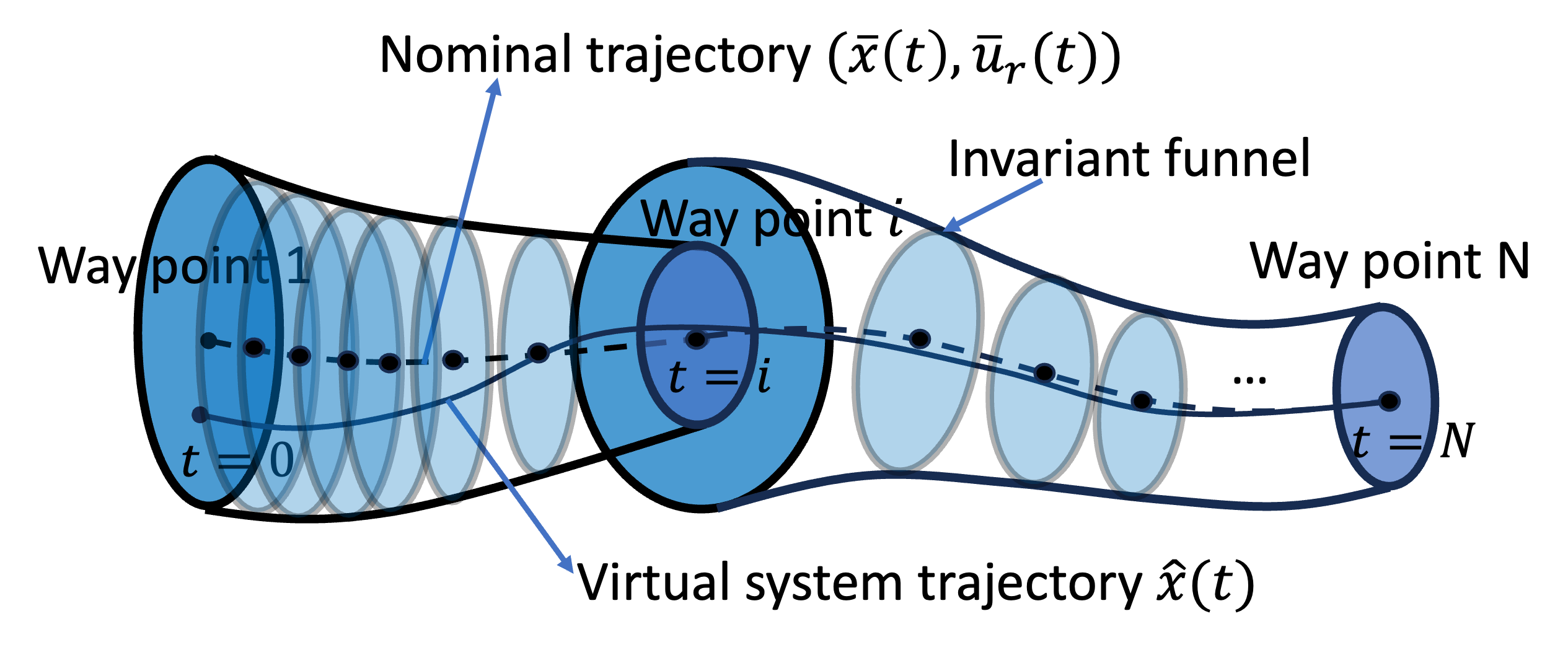}
    \caption{The architecture of the invariant funnels, where the funnels are constructed at each way point and the virtual system states $\hat{x}(t)$ are close to the nominal state trajectory $\bar{x}(t)$, for all $t\in\mathcal{I}_{N}^0$. The last way point at $t=N$ is the desired equilibrium point. }
    \label{fig:invariant_funnels}
    \vspace{-1mm}
\end{figure}

The basic architecture of the invariant funnels is shown in Fig. \ref{fig:invariant_funnels}, where we require that any states within the invariant funnels will never escape from there. Besides, if at the first way point, our states stay within the funnel created by the designed controller $K_{0}$, then at the next time step, the states will also stay within the funnel created by the controller $K_{1}$ at this way point.

To achieve funnel-based recovery control synthesis, in this paper, we define the specific class of invariant funnels as the following ellipsoidal space, i.e.,
\begin{align}\label{Eq:state_funnel}
    \mathcal{E}_{\eta_t}:=\Big\{\eta_t\in\mathbb{R}^n\ \Big|\ \eta_t^\T P_t^{-1}\eta_t\leq \frac{1}{r_t}\Big\},
\end{align}
where the constant $\frac{1}{r_t}$ defines the radius of the ellipsoid with $r_t\in\mathbb{R}_+$, for all $t\in\mathcal{I}_N^0$.
$P_t\in\mathbb{R}_+^{n\times n}$ is the matrix obtained by the DMI \eqref{Eq:DMI_main_condition} at the time step $t$. 
With the time-varying $P_t$ and $r_t$, the funnels constructed by \eqref{Eq:state_funnel} is invariant for any solution $\eta(\cdot)$ of \eqref{Eq:tracking_error_closed-loop_Lur'e_dynamics}, if $\eta_0\in\mathcal{E}_{\eta_0}$, then $\eta_t\in\mathcal{E}_{\eta_t}$, for all $t\in\mathcal{I}_{N}^0$.

For funnel-based design, apart from control synthesis using the DMI in \eqref{Eq:DMI_main_condition}, we also need to find the maximum invariant funnels for states and control inputs of the virtual system \eqref{Eq:virtual_system_learned_unknown_dynamics} within feasible state and input spaces $\hat{\mathcal{X}}$ and $\hat{\mathcal{U}}$, respectively.
% $:=\{\hat{x}\in\mathbb{R}^n\ |\ h_i(\hat{x})\leq 0,\ i\in\mathcal{I}_{l_x}\}$ and $\hat{\mathcal{U}}:=\{\hat{u}_r\in\mathbb{R}^m\ |\ g_j(\hat{u}_r)\leq 0,\ j\in\mathcal{I}_{l_u}\}$, respectively, where all the functions $h_i$ and $g_j$ are concave and differentiable at least once, for all $i\in\mathcal{I}_{l_x}$ and $j\in\mathcal{I}_{l_u}$. 
In this paper, we use linear polytopes constructed by several planar surfaces to approximate these two spaces, which are obtained by linearization around the nominal trajectories (way points) $\{\bar{x}_t\}_{t=0}^{N}$ and $\{\bar{u}_t\}_{t=0}^{N-1}$, respectively, as follows \cite{kim2024joint}:
\begin{subequations}\label{Eq:state_input_polytopes}
    \begin{align}
        \mathcal{P}_{\hat{x}_t}=\{x\ | (a_{it}^x)^\T \hat{x}_t\leq b_{it}^x,\ i\in\mathcal{I}_{l_x}\},\ \forall  t\in\mathcal{I}_{N}^0   \label{Eq:state_polytope}     \\
        \mathcal{P}_{\hat{u}_{rt}}=\{u\ | (a_{jt}^u)^\T \hat{u}_{rt}\leq b_{jt}^u,\ j\in\mathcal{I}_{l_u}\},\ \forall t\in\mathcal{I}_{N-1}^0.     \label{Eq:input_polytope}
    \end{align}
\end{subequations}
In this case, the inclusions $\mathcal{P}_{\hat{x}_t}\subseteq\mathcal{X}$ and $\mathcal{P}_{\hat{u}_{rt}}\subseteq\mathcal{U}$ hold, since the functions $h_i$ and $g_j$ are concave, for all $i\in\mathcal{I}_{l_x}$ and $j\in\mathcal{I}_{l_u}$.

Then, the following lemma can be established to ensure the maximal volume ellipsoid within a linear polytope region.

\begin{lemma}\label{Lem:state_state_constraints+ellipsoids}
    With $P_t$ solved by the DMI in \eqref{Eq:DMI_main_condition}, for the state tracking error $\eta_t:=\hat{x}_t-\bar{x}_t$ in \eqref{Eq:tracking_error_Lur'e_dynamics}, where the states $\hat{x}_t\in\mathcal{P}_{\hat{x}_t}$, for all $t\in\mathcal{I}_N^0$,
    % are within the feasible region defined by a linear polytope $\mathcal{P}_{x_t}:=\{x_t\in\mathbb{R}^n\ |\ a_i^\T x_t\leq b_i,\ i\in\mathcal{I}_{l_x}\}$, for all $t\in\mathcal{I}_N^0$, 
    the following condition
    \begin{align}\label{Eq:state_funnel+state_constraints}
        \bmtx{(b_{it}^x-(a_{it}^x)^\T\bar{x}_t)^2 r_t & (a_{it}^x)^\T P_t \\ P_t a_{it}^x  & P_t}\geq 0,\ \forall i\in\mathcal{I}_{l_x},
    \end{align}
    ensures the maximal volume ellipsoid $\max_{\eta_t}\{\mathcal{E}_{\eta_t}\}$ \eqref{Eq:state_funnel} is within the linear polytope $\mathcal{P}_{\hat{x}_t}$, i.e., $\bar{x}_t\oplus\max_{\eta_t}\{\mathcal{E}_{\eta_t}\}\subseteq \mathcal{P}_{\hat{x}_t}$, for all $t\in\mathcal{I}_N^0$.
\end{lemma}
\begin{proof}
    For all state tracking errors $\eta_t:=\hat{x}_t-\bar{x}_t$ satisfying the ellipsoidal constraint in \eqref{Eq:state_funnel}, where the states $\hat{x}_t$ in this error stay within the polytope $\mathcal{P}_{\hat{x}_t}$, we have:
    \begin{align}\label{Eq:state_constraints+ellipsoids}
        & (a_{it}^x)^\T(\bar{x}_t+\eta_t)\leq b_{it}^x,\ \ \forall\eta_t\in\mathcal{E}_{\eta_t},    \nonumber \\
        \Leftrightarrow\ & (a_{it}^x)^\T\bar{x}_t+\max_{\eta_t\in\mathcal{E}_{\eta_t}} (a_{it}^x)^\T\eta_t\leq b_{it}^x,    
    \end{align}
    for all $t\in\mathcal{I}_N^0$ and $i\in\mathcal{I}_{l_x}$.

    Define  $z_t:=P_t^{-\frac{1}{2}}\eta_t$. Then, $\eta_t=P_t^{\frac{1}{2}}z_t$ and the ellipsoidal constraint in \eqref{Eq:state_funnel} can be equivalently written as $\mathcal{E}_{z_t}:=\{z_t\in\mathbb{R}^n\ |\ z_t^\T z_t\leq \frac{1}{r_t}\}$. Consequently,
    \begin{align}
        \max_{\eta_t\in\mathcal{E}_{\eta_t}} (a_{it}^x)^\T\eta_t=\max_{z_t\in\mathcal{E}_{z_t}}(a_{it}^x)^\T P_t^{\frac{1}{2}}z_t,
    \end{align}
    for all $t\in\mathcal{I}_N^0$ and $i\in\mathcal{I}_{l_x}$.

    By applying the Cauchy-Schwarz inequality, we have:
    \begin{align}\label{Eq:Cauchy-Schwarz_condition}
        \max_{z_t\in\mathcal{E}_{z_t}}(a_{it}^x)^\T P_t^{\frac{1}{2}}z_t\leq \max_{z_t\in\mathcal{E}_{z_t}}|P_t^{\frac{1}{2}}a_{it}^x| |z_t|,
    \end{align}
    for all $t\in\mathcal{I}_N^0$ and $i\in\mathcal{I}_{l_x}$, where the equality condition holds when the vectors $P_t^{\frac{1}{2}}a_{it}^x$ and $z_t$ are parallel, i.e., $z_t=\mu_t P_t^{\frac{1}{2}}a_{it}^x$ with some $\mu_t\in\mathbb{R}$, in which $\mu_t=\frac{1}{\sqrt{r_t}|P_t^{\frac{1}{2}}a_{it}^x|}$, since $\max_{z_t} z_t^\T z_t=\frac{1}{r_t}$.
    
    Therefore, from \eqref{Eq:Cauchy-Schwarz_condition}, we have:
    \begin{align}\label{Eq:maximum_aT_P_a}
        \mbox{LHS of } \eqref{Eq:Cauchy-Schwarz_condition}&=\max_{z_t\in\mathcal{E}_{z_t}}|P_t^{\frac{1}{2}}a_{it}^x| |z_t|    \nonumber  \\
        &=\frac{1}{\sqrt{r_t}}|P_t^{\frac{1}{2}}a_{it}^x|=\frac{1}{\sqrt{r_t}}\sqrt{(a_{it}^x)^\T P_t a_{it}^x},
    \end{align}
    for all $t\in\mathcal{I}_N^0$ and $i\in\mathcal{I}_{l_x}$. 
    
    Substituting this into \eqref{Eq:state_constraints+ellipsoids}, we have:
    \begin{align}
        & (a_{it}^x)^\T\bar{x}_t+\frac{1}{\sqrt{r_t}}\sqrt{(a_{it}^x)^\T P_t a_{it}^x}\leq b_{it}^x,  \nonumber  \\
        \Leftrightarrow\ \ & \frac{1}{\sqrt{r_t}}\sqrt{(a_{it}^x)^\T P_t a_{it}^x}\leq b_{it}^x-(a_{it}^x)^\T\bar{x}_t,
    \end{align}
    for all $t\in\mathcal{I}_N^0$ and $i\in\mathcal{I}_{l_x}$. 
    Then, squaring both sides and using Schur complement, the above condition is equivalent to that in \eqref{Eq:state_funnel+state_constraints}. This completes the proof.
\end{proof}

Similar to the above state constraints, if there exist any input constraints, we can also construct invariant funnels of the control input for the designed controller \eqref{Eq:full_state_feedback_controller} using the following results.
\begin{corollary}\label{Cor:control_input_ellipsoid}
    With $P_t$ and $K_t$ satisfying the DMI in \eqref{Eq:DMI_main_condition} and the state funnel \eqref{Eq:state_funnel}, the following ellipsoid 
    \begin{align}\label{Eq:input_funnel}
        \mathcal{E}_{\zeta_{t}}:=\{(K_tP_tK_t^\T)^{\frac{1}{2}}\zeta_t\ |\ |\zeta_t|\leq \frac{1}{\sqrt{r_t}},\ \zeta_t\in\mathbb{R}^m\}
    \end{align} 
    is the invariant funnels of control input $\xi_t$ in \eqref{Eq:full_state_feedback_controller}, for all $t\in\mathcal{I}_{N-1}^0$. 
    Correspondingly, the following control input constraint
    \begin{align}\label{Eq:input_funnel+input_constraints}
        \bmtx{(b_{jt}^u-(a_{jt}^u)^\T\bar{u}_{rt})^2 r_t & (a_{jt}^u)^\T K_t P_t \\ (\star)^\T  & P_t}\geq 0,
    \end{align}
    for all $j\in\mathcal{I}_{l_u}$, ensures the maximal volume ellipsoid $\max_{y_t}\{\mathcal{E}_{y_{t}}\}$ within the input linear polytope $\mathcal{P}_{\hat{u}_{rt}}$, i.e., $\bar{u}_{rt}\oplus\max_{y_t}\{\mathcal{E}_{y_{t}}\}\subseteq\mathcal{P}_{\hat{u}_{rt}}$, for all $t\in\mathcal{I}_{N-1}^0$.
\end{corollary}
\begin{proof}    
    Similar to the proof of Lem. \ref{Lem:state_state_constraints+ellipsoids}, we still use the state transformation $z_t:=P_t^{-\frac{1}{2}}\eta_t\in\mathbb{R}^n$, for all $t\in\mathcal{I}_{N-1}^0$. 
    Then, the ellipsoid of the control input is $\mathcal{E}_{\xi_t}=\{K_t\eta_t\ |\ \eta_t^\T P_t^{-1}\eta_t\leq\frac{1}{r_t}\}$, for all $t\in\mathcal{I}_{N-1}^0$. 
    
    With the transformed state $z_t$, the control input $\xi_t$ in \eqref{Eq:full_state_feedback_controller} can be rewritten as:
    \begin{align}
        \xi_t=K_t\eta_t=M_t z_t,
    \end{align}
    where $M_t:=K_tP_t^{\frac{1}{2}}$. 
    % Thus, the ellipsoid $\mathcal{E}_{\xi_t}$ for the control input designed by \eqref{Eq:full_state_feedback_controller} can be written as $\mathcal{E}_{\xi_t}=\{\xi_t=M_tz_t\ |\ |z_t|^2\leq\frac{1}{r_t}\}$, for all $t\in\mathcal{I}_{N-1}^0$. 

    Using singular value decomposition (SVD) of $M_t$, we have:
    \begin{align}\label{Eq:Mt_SVD}
        M_t=U_{rt}\Sigma_{rt}V_{rt}^\T,
    \end{align}
    where $c_t=\rank(M_t)$, $U_{rt}\in\mathbb{R}^{m\times c_t}$ and $V_{rt}\in\mathbb{R}^{n\times c_t}$ are orthonormal matrices, and $\Sigma_{rt}\in\mathbb{R}^{c_t\times c_t}$ is a positive diagonal matrix.

    To prove \eqref{Eq:input_funnel}, we consider two sets $\mathcal{A}:=\{M_t z_t\in\mathbb{R}^m\ |\ |z_t|^2\leq\frac{1}{r_t},\ z_t\in\mathbb{R}^n\}$ and $\mathcal{B}:=\{(M_tM_t^\T)^{\frac{1}{2}}\zeta_t\ |\ |\zeta_t|^2\leq\frac{1}{r_t},\ \zeta_t\in\mathbb{R}^m\}$. Hence, it is suffice to show $\mathcal{A}=\mathcal{B}$.
    To show this, we discuss both sets separately.
    % Then, $M_tM_t^\T=U_{rt}\Sigma_{rt}^2U_{rt}^\T$ and $(M_tM_t^\T)^{\frac{1}{2}}=U_{rt}\Sigma_{rt}U_{rt}^\T$.

    1) For the set $\mathcal{A}$, take any $z_t\in\mathbb{R}^n$ with $|z_t|^2\leq\frac{1}{r_t}$. Note that $z_t$ can be written as:
    \begin{align}
        z_t = V_{rt}y_t+z_{t}^{\perp},
    \end{align}
    where $y_t:=V_{rt}^\T z_t\in\mathbb{R}^{c_t}$ and $z_t^{\perp}\in\ker(V_{rt}^\T)$, i.e., $V_{rt}^\T z_t^{\perp}=\0$. Since the columns of $V_{rt}$ are orthonormal, the condition $|y_t|^2\leq |z_t|^2\leq\frac{1}{r_t}$ is satisfied. 
    
    Now, we have:
    \begin{align}
        M_t z_t=U_{rt}\Sigma_{rt} V_{rt}^\T z_t=U_{rt}\Sigma_{rt}y_t,
    \end{align}
    and thus, every element in the set $\mathcal{A}$ is $U_{rt}\Sigma_{rt}y_t$ for some $y_t\in\mathbb{R}^{c_t}$ with $|y_t|^2\leq\frac{1}{r_t}$. In other words, the set $\mathcal{A}$ is equivalent to the set $\{U_{rt}\Sigma_{rt}y_t\ |\ |y_t|^2\leq\frac{1}{r_t},\ y_t\in\mathbb{R}^{c_t}\}$.

    2) For the set $\mathcal{B}$, take any $\zeta_t\in\mathbb{R}^m$ with $|\zeta_t|^2\leq\frac{1}{r_t}$. Let $\omega_t:=U_{rt}^\T\zeta_t\in\mathbb{R}^{c_t}$. Then, $|\omega_t|^2\leq|\zeta_t|^2\leq\frac{1}{r_t}$ is satisfied, since $U_{rt}$ has orthonormal columns. 
    
    With the $M_t$ matrix in \eqref{Eq:Mt_SVD}, we have:
    \begin{align}
        (M_tM_t^\T)^{\frac{1}{2}}\zeta_t=U_{rt}\Sigma_{rt}U_{rt}^\T\zeta_t=U_{rt}\Sigma_{rt}\omega_t,
    \end{align}
    and thus, every element in the set $\mathcal{B}$ is $U_{rt}\Sigma_{rt}\omega_t$ for some $\omega_t\in\mathbb{R}^{c_t}$ with $|\omega_t|^2\leq\frac{1}{r_t}$. Therefore, the set $\mathcal{B}$ is equivalent to the set $\{U_{rt}\Sigma_{rt}\omega_t\ |\ |\omega_t|^2\leq\frac{1}{r_t},\ \omega_t\in\mathbb{R}^{c_t}\}$.

    Combining the above cases 1) and 2), it is readily shown that the sets $\mathcal{A}=\mathcal{B}$. This completes the proof of the selection of control input ellipsoid \eqref{Eq:input_funnel}.

    With the ellipsoidal space of control input \eqref{Eq:input_funnel}, the proof of the input constraint \eqref{Eq:input_funnel+input_constraints} follows a similar procedure as our proof of the state constraint in Thm. \ref{Th:invariant_funnels_condition}, which can be readily shown and omitted here. This completes the proof. 
\end{proof}

% When $M_tM_t^\T=K_tP_tK_t^\T$ is invertible, we have:    
%     \begin{align}
%         \xi_t^\T(M_tM_t^\T)^{-1}\xi_t=z_t^\T M_t^\T(M_tM_t^\T)^{-1}M_tz_t\leq \frac{1}{r_t}.
%     \end{align}
%     Then, take variable transformation as $y_t:=(K_tP_tK_t^\T)^{-\frac{1}{2}}\eta_t$, for all $t\in\mathcal{I}_{N}^0$.

With the ellipsoid in \eqref{Eq:state_funnel}, we have the following theorem for the invariant funnel estimation along the nominal trajectory under the designed controller $K_t$ by solving \eqref{Eq:DMI_main_condition}. 
In other words, we aim to find the invariant funnels for states $\eta_t$ in \eqref{Eq:tracking_error_closed-loop_Lur'e_dynamics}, such that for any solutions $\eta(\cdot)$ of \eqref{Eq:tracking_error_closed-loop_Lur'e_dynamics}, if $\eta_{t}\in\mathcal{E}_{\eta_t}$, then $\eta_{t+1}\in\mathcal{E}_{\eta_{t+1}}$, for all $t\in\mathcal{I}_{N-1}^0$.

\begin{theorem}\label{Th:invariant_funnels_condition}
    With $P_t$, $K_t$ and $\gamma_2$ satisfying the DMI in \eqref{Eq:DMI_main_condition} and the state and input constraints in \eqref{Eq:state_funnel+state_constraints} and \eqref{Eq:input_funnel+input_constraints}, the following ellipsoid
    \begin{align}\label{Eq:invariant_ellipsoidal_funnels}
        \mathcal{E}_{\eta_t}:=\Big\{\eta_t\in\mathbb{R}^n\ \Big|\ V_t:=\eta_t^\T P_t^{-1}\eta_t\leq \frac{1}{r_t} \Big\}
    \end{align}
    is invariant for all $t\in\mathcal{I}_{N}^0$, 
    where $r_t$ is selected by:
    \begin{align}\label{Eq:funnel_size_r_t}
        r_t^{-1} \geq 
        \begin{cases}
        \gamma_2\|\hat{\Delta}(\cdot)\|_{\infty}^2,\ & k_t\leq 0,  \\
        \frac{\gamma_2}{1-k_{\max}}\|\hat{\Delta}(\cdot)\|_{\infty}^2,\ & k_t\in(0,1),
    \end{cases}     
    \end{align}
    with $k_t=1-\frac{1}{\gamma_{2}\lambda_t^{\max}}$, $k_{\max}$ is the maximum value of $k_t$ over all $t\in\mathcal{I}_N^0$, and $\lambda_t^{\max}$ is the maximum eigenvalue of the matrix $P_t^{-1}$ at the time step $t$, for all $t\in\mathcal{I}_{N}^0$.
\end{theorem}

\begin{proof}
    % Now, we need to show that the satisfaction of \eqref{Eq:DMI_main_condition} ensures the invariance property of the state and input funnels. 
    Note that for the condition \eqref{Eq:V_dot2} in the proof of Thm. \ref{Th:L2_stability_main_condition}, we have:
\begin{align}\label{Eq:rewritten_condition}
    & \Delta V_t +\lambda_{pt}(\gamma_{1t}^2|\tilde{q}_t|^2-|\tilde{p}_t|^2) \leq \gamma_{2}|\hat{\Delta}_t|^2-\frac{1}{\gamma_{2}}|\eta_t|^2,    \nonumber   \\
    \Rightarrow\ & V_{t+1}-\alpha V_t \leq \gamma_{2}|\hat{\Delta}_t|^2-\frac{1}{\gamma_{2}}|\eta_t|^2,\ (\alpha\in(0,1])    \nonumber  \\
    \Rightarrow\ & V_{t+1}-V_t\leq \Delta V_t \leq \gamma_{2}|\hat{\Delta}_t|^2-\frac{1}{\gamma_{2}}|\eta_t|^2.
\end{align}

For the Lyapunov function in \eqref{Eq:time-varying_Lyapunov_function}, we have:
\begin{align}\label{Eq:Lyapunov_relation}
    \lambda_{\min}(P_t^{-1})|\eta_t|^2\leq\eta_t^\T P_t^{-1}\eta_t\leq \lambda_{\max}(P_t^{-1})|\eta_t|^2,
\end{align}
where $\lambda_{\min}(P_t^{-1})$ and $\lambda_{\max}(P_t^{-1})$ are the minimum and maximum eigenvalues of the matrix $P_t^{-1}$, for all $t\in\mathcal{I}_{N}^0$. We denote these two eigenvalues as $\lambda_t^{\min}$ and $\lambda_t^{\max}$, respectively.

Using \eqref{Eq:Lyapunov_relation} in \eqref{Eq:rewritten_condition}, we have:
\begin{align}\label{Eq:bound_Vt+1}
    & V_{t+1}-V_t\leq\gamma_{2}|\hat{\Delta}_t|^2-\frac{1}{\gamma_{2}\lambda_t^{\max}}V_t,   \nonumber     \\
    \Leftrightarrow\ \ & V_{t+1}\leq \gamma_{2}|\hat{\Delta}_t|^2+\Big(1-\frac{1}{\gamma_{2}\lambda_t^{\max}}\Big) V_t,
\end{align}
for all $t\in\mathcal{I}_{N-1}^0$.

% Then, to find the invariant funnels for the closed-loop system \eqref{Eq:tracking_error_closed-loop_Lur'e_dynamics} under the designed controller, if the initial funnel is satisfied with $V_{0}\leq r_0^{-1}$, where $r_0\in\mathbb{R}_+$, we require the following condition to hold from \eqref{Eq:bound_Vt+1}:
% \begin{align}\label{Eq:bound_Vt+1_rt+1}
%    V_{t+1}\leq r_{t+1}^{-1}\leq \gamma_{2}|\hat{\Delta}_t|^2+\Big(1-\frac{1}{\gamma_{2}\lambda_t^{\max}}\Big) V_t,
% \end{align}
%for $r_{t+1}\in\mathbb{R}_+$, for all $t\in\mathcal{I}_{N-1}^0$, since this is a sub-level constraint on $V_t$, for all $t\in\mathcal{I}_{N}^0$. 
% Note that if the condition \eqref{Eq:bound_Vt+1_rt+1} is invariant and satisfied, then the condition \eqref{Eq:bound_Vt+1}, i.e., the condition \eqref{Eq:bound_Vt+1_rt+1} with the smallest $r_t$ values, is also invariant, for all $t\in\mathcal{I}_{N}^0$. 
% This implies that $V_t\leq r_t$ is always satisfied, for all for all $t\in\mathcal{I}_{N}^0$. 

Define $k_t:=1-\frac{1}{\gamma_{2}\lambda_t^{\max}}$ in \eqref{Eq:bound_Vt+1}, for all $t\in\mathcal{I}_N^0$. 
If at certain time step, the following condition is satisfied:
\begin{align}
   V_t\leq r_t^{-1}, 
\end{align}
where $r_t\in\mathbb{R}_+$ satisfying \eqref{Eq:funnel_size_r_t}, for all $t\in\mathcal{I}_{N-1}^0$. 
To discuss the conditions in the next time step, due to the fact that $\gamma_2$, $\lambda_t^{\max}\in\mathbb{R}_+$, there are two possible cases for $V_{t+1}$ based on different values of $k_t$ in \eqref{Eq:bound_Vt+1}, i.e., 1) $k_t\leq 0$, and 2) $k_t\in(0,1)$.

Thus, for case 1) in the next time step, we directly have from \eqref{Eq:bound_Vt+1} with the selected funnel (the first condition in \eqref{Eq:funnel_size_r_t}):
\begin{align}
    V_{t+1}\leq\gamma_{2}|\hat{\Delta}_t|^2\leq\gamma_{2}\|\hat{\Delta}(\cdot)\|_{\infty}^2\leq r_{t+1}^{-1},
\end{align}
for all $t\in\mathcal{I}_{N-1}^0$, since the second term on the right-hand side of \eqref{Eq:bound_Vt+1} is non-positive. Therefore, in case 1), the ellipsoid $V_{t}\leq\gamma_{2}\|\hat{\Delta}(\cdot)\|_{\infty}^2\leq r_t^{-1}$ is invariant for all $t\in\mathcal{I}_{N}^0$. 
In other words, for any solutions $\eta(\cdot)$ of \eqref{Eq:tracking_error_closed-loop_Lur'e_dynamics}, if $\eta_{t}\in\mathcal{E}_{\eta_t}$, then $\eta_{t+1}\in\mathcal{E}_{\eta_{t+1}}$, for all $t\in\mathcal{I}_{N-1}^0$.

In case 2), assume that there exists a constant $c\in\mathbb{R}_+$ such that $V_t\leq c\leq r_t^{-1}$ at certain time step $t\in\mathcal{I}_{N-1}^0$.
For the condition \eqref{Eq:bound_Vt+1} in case 2) at the next time step, we require:
\begin{align}
    & V_{t+1}\leq \sup_{t}\{\gamma_{2}|\hat{\Delta}_t|^2+k_t c\}= c,\ \forall t\in\mathcal{I}_{N-1}^0,    \nonumber \\
    \hspace{-2mm}\Rightarrow\ & c= \frac{\gamma_2}{1-k_{\max}}\|\hat{\Delta}(\cdot)\|_{\infty}^2,\ k_{\max}:=\max_t\{k_t\}\in(0,1), \nonumber  \\
    \Rightarrow\ & V_{t+1}\leq c\leq r_{t+1}^{-1},\ (\mbox{by the selction in }\eqref{Eq:funnel_size_r_t}).
\end{align}
Therefore, in case 2), the bound $V_{t}\leq\frac{\gamma_2}{1-k_{\max}}\|\hat{\Delta}(\cdot)\|_{\infty}^2\leq r_t^{-1}$ (i.e., the second condition in \eqref{Eq:funnel_size_r_t}) is invariant, for all $t\in\mathcal{I}_{N}^0$.

Therefore, combining the above two cases, we have shown that the following bound:
\begin{align}
    V_t &\leq r_t^{-1},
\end{align}
is invariant, where $r_t^{-1}\geq\begin{cases}
        \gamma_2\|\hat{\Delta}(\cdot)\|_{\infty}^2,\ & k_t\leq 0  \\
        \frac{\gamma_2}{1-k_{\max}}\|\hat{\Delta}(\cdot)\|_{\infty}^2,\ & k_t\in(0,1)
    \end{cases}$,
for all $t\in\mathcal{I}_{N}^0$. This completes the proof.
\end{proof}

% \begin{remark}
%     In Thm. \ref{Th:invariant_funnels_condition}, when $k\in(0,1)$, we can select any $c\geq\frac{\gamma_2}{1-k}\|\hat{\Delta}(\cdot)\|_{\infty}^2$ and the resulting ellipsoid $V_t\leq c$ is still invariant. However, arbitrarily large $c$ values are trivial solutions of the invariant ellipsoid, which does not provide the ability to resist the impact of unknown dynamics.
% \end{remark}

\begin{remark}
    In Thm. \ref{Th:invariant_funnels_condition}, the values of $r_t^{-1}$ can be selected satisfying the two cases in \eqref{Eq:funnel_size_r_t}, and the resulting ellipsoids (i.e., the funnels) are still invariant. However, arbitrarily large $r_t^{-1}$ values are trivial solutions of the invariant funnel, which does not provide the ability to resist the impact of unknown dynamics.
\end{remark}

Now, with the invariant funnels obtained in Thm. \ref{Th:invariant_funnels_condition} and Schur complement, the funnel-based control synthesis problem can be formulated as:
\vspace{-1mm}
\begin{subequations}\label{Eq:Th:discrete-time_funnel_synthesis}
        \begin{align}
        \min_{\substack{P_t, \mu_t^P, \mu_t^K, \\
        Y_t, \nu_t}}&  w_1\sum_{t=0}^N\mu_t^P+w_2 \sum_{t=0}^{N-1}\mu_t^K-w_3\log\det(P_0)    \\
        \mbox{s.t. }\ \ \ & 0< P_t\leq \mu_t^P\I,\ \forall t\in\mathcal{I}_{N}^0,     \\ 
        & \bmtx{\mu_t^K\I & Y_t \\ Y_t^\T & P_t}\geq 0,\ \forall t\in\mathcal{I}_{N-1}^0,    \\
        & \eqref{Eq:funnel_size_r_t},\ \eqref{Eq:DMI_main_condition},\ \eqref{Eq:state_funnel+state_constraints}, \mbox{ and } \eqref{Eq:input_funnel+input_constraints},    \\
        & P_0\geq r_0 P_i,\ P_N\leq r_f P_f,    \label{Eq:initial_final_P_constraints}
    \end{align}
\end{subequations}
where the $L_2$-gain $\gamma_2$ in \eqref{Eq:DMI_main_condition} is pre-selected according to the application scenarios.
$w_1$, $w_2$, $w_3\in\mathbb{R}_+$ are pre-defined parameters.
$P_i$, $P_f\in\mathbb{R}_+^{n\times n}$ are pre-defined initial and final ellipsoidal matrices, the controller gain $K_t$ in \eqref{Eq:full_state_feedback_controller} is determined by $K_t=Y_tP_t^{-1}$, for all $t\in\mathcal{I}_{N-1}^0$.

Note that the invariant funnel size $r_t^{-1}$ used in \eqref{Eq:state_funnel+state_constraints}, \eqref{Eq:input_funnel+input_constraints} and \eqref{Eq:initial_final_P_constraints} are non-convex even with a given $L_2$-gain $\gamma_2$.
For the funnel size $r_t^{-1}$, the condition when $k_t\leq 0$ is convex for the given $\gamma_2$ and $\|\hat{\Delta}(\cdot)\|_{\infty}^2$ values, while for $k_t\in(0,1)$, we notice that the $k_{t}$ in \eqref{Eq:funnel_size_r_t} can be equivalently written as:
\begin{align}
    k_t=1-\frac{1}{\gamma_{2}\lambda_{\max}(P_t^{-1})} = 1-\frac{\lambda_{\min}(P_t)}{\gamma_2},
\end{align}
which is convex for $\lambda_{\min}(P_t)$ with a given $L_2$-gain value.

Using this property, we can modify the funnel condition \eqref{Eq:funnel_size_r_t} into soft constraints and reformulate the problem in \eqref{Eq:Th:discrete-time_funnel_synthesis} for online synthesis as follows:
\vspace{-2mm}
\begin{subequations}\label{Eq:Th:discrete-time_online_funnel_synthesis}
        \begin{align}
        \min_{\substack{P_t, \mu_t^P, \mu_t^K, \\
        Y_t, \nu_t, r_t}}&  w_1\sum_{t=t_1}^{t_2}\mu_{2t}^P+w_2 \sum_{t=t_1}^{t_2-1}\mu_t^K-w_3\log\det(P_{t_1})+  \nonumber     \\ 
        & w_4 \mu_1^P-w_5\sum_{t=t_1}^{t_2} r_t       \\
        \mbox{s.t. }\ \ \ &  0 \leq r_t \leq \bar{r},\ 0 \leq r_t \leq \bar{r}(1-k_t),\ t=t_1,...,t_2,  \label{Eq:relaxed_funnel_size_constraints} \\
        & 0 < \mu_1^P \I \leq P_t\leq \mu_{2t}^P\I,\ t=t_1,...,t_2,     \\ 
        & \bmtx{\mu_t^K\I & Y_t \\ Y_t^\T & P_t}\geq 0,\ t=t_1,,...,t_2-1    \\
        & \eqref{Eq:DMI_main_condition}, \eqref{Eq:input_funnel+input_constraints},\ t = t_1,...,t_2-1 \\
        & \eqref{Eq:state_funnel+state_constraints},\ t=t_1,...,t_2,    \\
        & P_{t_1}\geq r_{t_1} P_i,\ P_{t_2}\leq r_{t_2} P_f,    \label{Eq:t1_t2_P_constraints}
    \end{align}
\end{subequations}
where $t_1$, $t_2\in\mathcal{I}_N^0$ formulate a sliding time window with $t_1<t_2$, $\bar{r}:=\frac{1}{\gamma_2\|\hat{\Delta}\|^2_{\infty}}$. 
$w_4$ and $w_5\in\mathbb{R}_+$ are pre-defined weights, and $w_5$ should be selected extremely large.
Here, in order to embed the funnel constraints in the commercial solvers (e.g., \texttt{cvxpy} in Python), we modify the (non-convex) binary constraints \eqref{Eq:funnel_size_r_t} into convex relaxations in \eqref{Eq:relaxed_funnel_size_constraints}. With the extremely large $w_5$ value, we can get as close as the ground truth value of $r_t$.

\begin{remark}
    For online control synthesis (solving problem \eqref{Eq:Th:discrete-time_online_funnel_synthesis}), the value of $\|\hat{\Delta}(\cdot)\|_{\infty}$ can be approximated by simply computing $|\hat{\Delta}_t|$, for all $t\in\mathcal{I}_N^0$, and the analysis in Thm. \ref{Th:invariant_funnels_condition} still holds. 
    Or conservatively, one may choose a relatively large constant estimation value of $\|\hat{\Delta}(\cdot)\|_{\infty}$. 
    Besides, the value of $k_{\max}$ can be locally approximated by the value of $k_{t}$, for all $t\in\mathcal{I}_{N}^0$.
    % since this will not violate the analysis in the proof of Thm. \ref{Th:invariant_funnels_condition}. 
    With the values $|\hat{\Delta}_t|$ and $k_t$ computed at the current time step, the funnel size $r_t^{-1}$ in \eqref{Eq:funnel_size_r_t} can be simply refined as:
    $r_t^{-1}\geq\begin{cases}
        \gamma_2|\hat{\Delta}_t|^2,\ & k_t\leq 0  \\
        \frac{\gamma_2}{1-k_{t}}|\hat{\Delta}_t|^2,\ & k_t\in(0,1)
    \end{cases}$, for all $t\in\mathcal{I}_N^0$, and this will not violate the analysis in the proof of Thm. \ref{Th:invariant_funnels_condition} for online designs. 
    While smaller $k_t$ values provide more conservative results (i.e., smaller minimal funnel size), they are allowed to be computed online and provide a tighter estimation for the bounds on tracking errors under disturbances.
\end{remark}

Based on the above discussions, we summarize our proposed recovery control design procedure in the following algorithm.
\vspace{-2mm}
\begin{algorithm}
\caption{Online Recovery Control and Funnel Synthesis for System \eqref{Eq:Lur'e_type_system}}
\label{alg:recovery_control_single_system_algorithm}
\begin{algorithmic}[1]
    \State \textbf{Input:} $Q$, $S$, $R$, initial conditions for REN, $\{\bar{x}_t\}_{t=0}^N$, $\{\bar{u}_t\}_{t=0}^{N-1}$, $\{\mathcal{P}_{\hat{x}_t}\}_{t=0}^N$, $\{\mathcal{P}_{\hat{u}_{rt}}\}_{t=0}^{N-1}$, $\{w_i:i\in\mathcal{I}_5\}$, $\alpha$, $\gamma_2$, $\|\hat{\Delta}\|_{\infty}$, initial conditions and system parameters for \eqref{Eq:Lur'e_type_system};
    \State \textbf{Output:} Real-time actual system trajectory $(x_t,u_r)$, real-time trajectory of the system with the learned unknown dynamics $(\hat{x}_t,\hat{u}_r)$, the learned unknown dynamics $\hat{\Delta}_t$ and the invariant funnels;
    \State Set the one step horizon of the REN training and funnel control;
    \For{each epoch in TotalEpochs}
        \For{$t$ in Horizon}
            \State Solve \eqref{Eq:Th:discrete-time_online_funnel_synthesis} to obtain the controller $K_t$;
            \State Train REN \eqref{Eq:state_transition_compact} with the guarantee of \eqref{Eq:incremental_IQC_REN};
            % and \eqref{Eq:incremental_IQC_REN_F};
            \State Generate the trajectory of actual dynamics \eqref{Eq:NonlinearSystemDynamics_withPriorControl} and the system with the learned unknown dynamics \eqref{Eq:Lur'e_type_system};
        
            \State Update the controller $K_t$ using the DMI \eqref{Eq:DMI_main_condition};
        
            \State Update the learned unknown dynamics $\hat{\Delta}$ by \eqref{Eq:state_transition_compact};
            \State Update the funnels \eqref{Eq:state_funnel+state_constraints} and \eqref{Eq:input_funnel+input_constraints}.
        \EndFor
    \EndFor
    \State \Return Real-time trajectories $(x_t,u_t)$, $(\hat{x}_t,\hat{u}_t)$, learned unknown dynamics $\hat{\Delta}_t$ and funnels size $r_t^{-1}$.
\end{algorithmic}
\end{algorithm}

%---------------------------------------------------------------
\section{Simulation Example}\label{sec:simulation}

\begin{figure}[!t]
    \centering
    \includegraphics[width=0.7\linewidth]{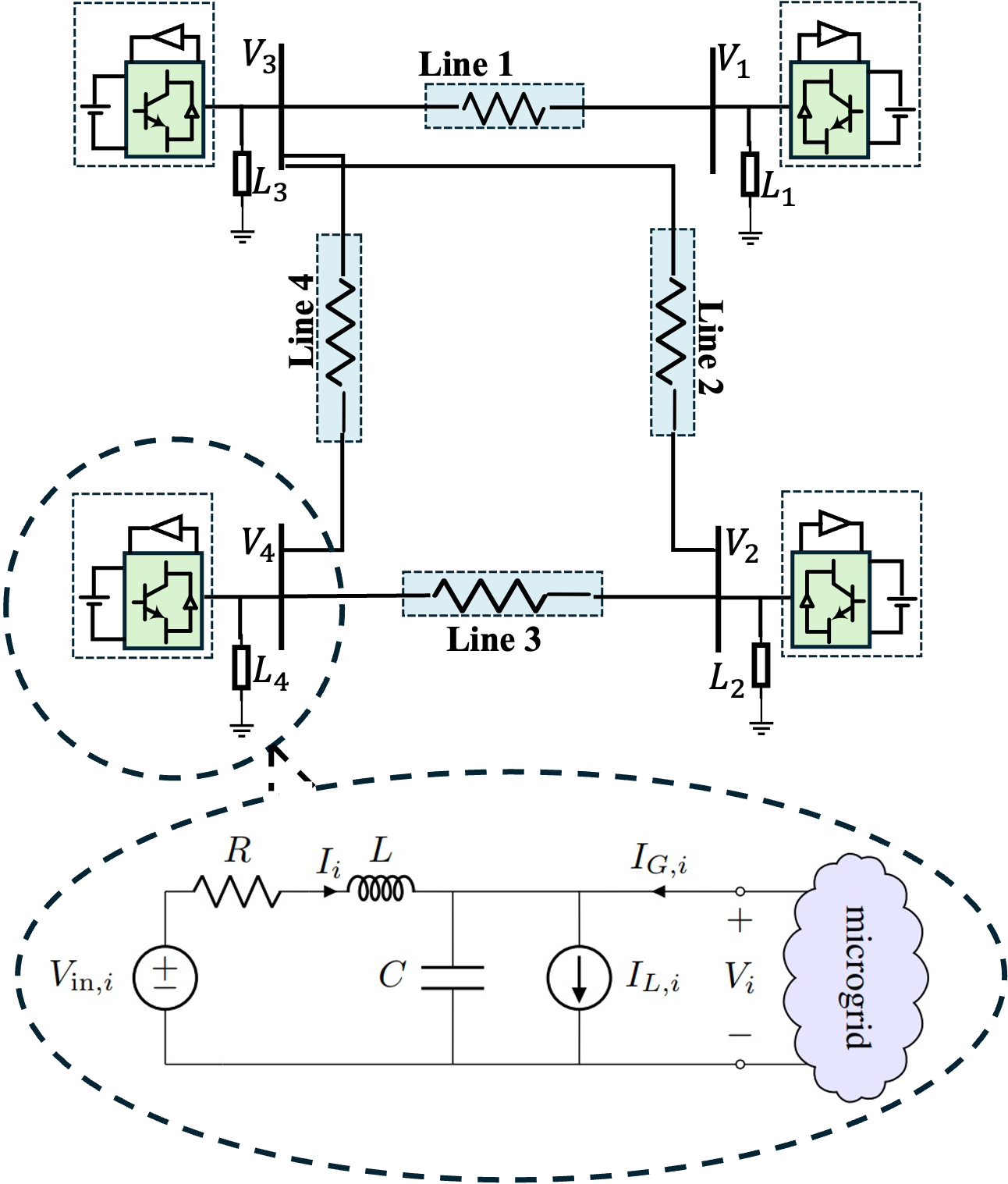}
    \caption{
    The architecture of our DC microgrid \cite{najafirad2025dissipativity,nakano2025dissipativity}.}
    \label{fig:DC_microgrid_architecture}
    \vspace{-1mm}
\end{figure}

In this section, we provide a simulation example to verify the effectiveness of our proposed methods.
In particular, we consider a DC microgrid network \cite{nakano2025dissipativity} that contains $4$ distributed generators (DGs) and $4$ power transmission lines, and the $i\tsup{th}$ subsystem dynamics is shown as follows:
\begin{align}\label{Eq:actual_DC_microgrid_subsystem}
    x_i(t+1) &= A_ix_i(t)+B_{1,i} u_i(t) + B_{2,i} v_i(t) + \Delta(t),  \\
    y_i(t) &= C_ix_i(t),
\end{align}
where $x_i=[V_i\ I_i]^\T$ is the system states, involving the DG voltage $V_i$ and current $I_i$, for all $i\in\mathcal{I}_4$. $u_i=V_{in,i}\in\mathbb{R}$ is the control input, which is also the DC source.
$v_i=\sum_{j\in\mathcal{N}_i} (y_j-y_i)$ represents the interconnections between the subsystems. The basic architecture of our considered DC microgrid is shown in Fig. \ref{fig:DC_microgrid_architecture}.

Other setups and system parameters of the systems are selected, comparable to those of \cite{nakano2025dissipativity}. 
Our goal of this example is to stabilize the internal current $I_i$ and the internal voltage $V_i$ of all DGUs with state-feedback controllers designed in a decentralized fashion on the basis of data. 
In this way, our proposed solutions are verified by a specifically developed simulator\footnote{Available at \href{https://github.com/NDzsong2/single_dynamics_funnel-main}{https://github.com/NDzsong2/single\textunderscore dynamics\textunderscore funnel-main.git}} constructed with TensorFlow and \texttt{cvxpy} solver.

\begin{figure}[!t]
    \centering
    \includegraphics[width=\linewidth]{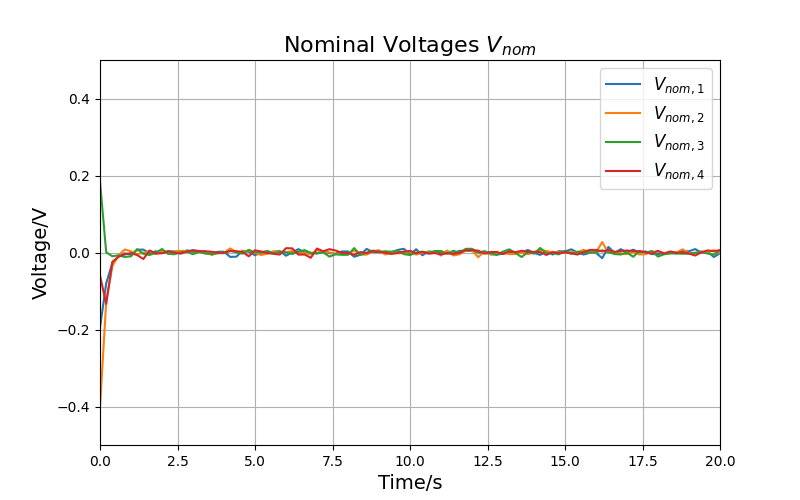}
    \caption{
    Nominal voltages.}
    \label{fig:nominal_voltages}
    \vspace{-1mm}
\end{figure}

\begin{figure}[!t]
    \centering
    \includegraphics[width=\linewidth]{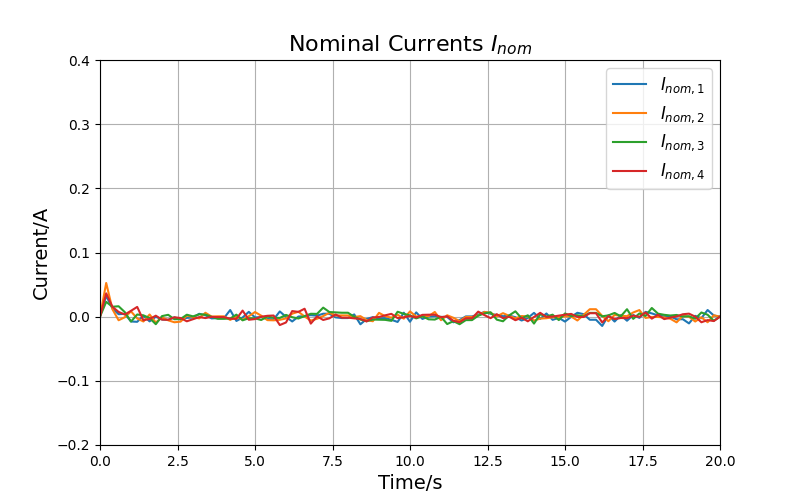}
    \caption{
    Nominal currents.}
    \label{fig:nominal_currents}
    \vspace{-1mm}
\end{figure}

\begin{figure}[!t]
    \centering
    \includegraphics[width=\linewidth]{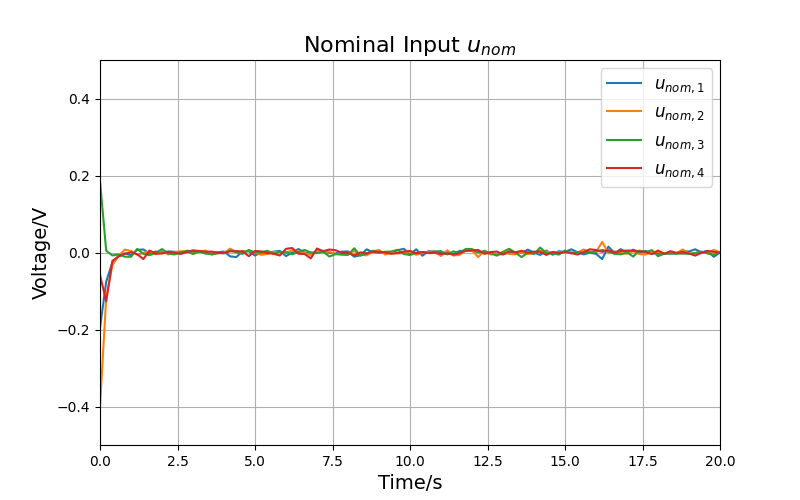}
    \caption{
    Nominal control inputs.}
    \label{fig:nominal_control_inputs}
    \vspace{-1mm}
\end{figure}

\begin{figure}[!t]
    \centering
    \includegraphics[width=\linewidth]{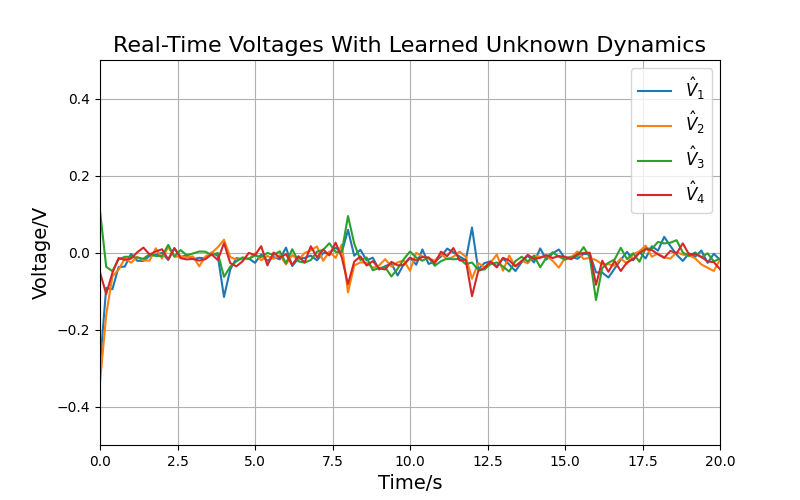}
    \caption{
    Real-time voltages for the system under the learned unknown dynamics.}
    \label{fig:real-time_voltages_estimated}
    \vspace{-1mm}
\end{figure}

\begin{figure}[!t]
    \centering
    \includegraphics[width=\linewidth]{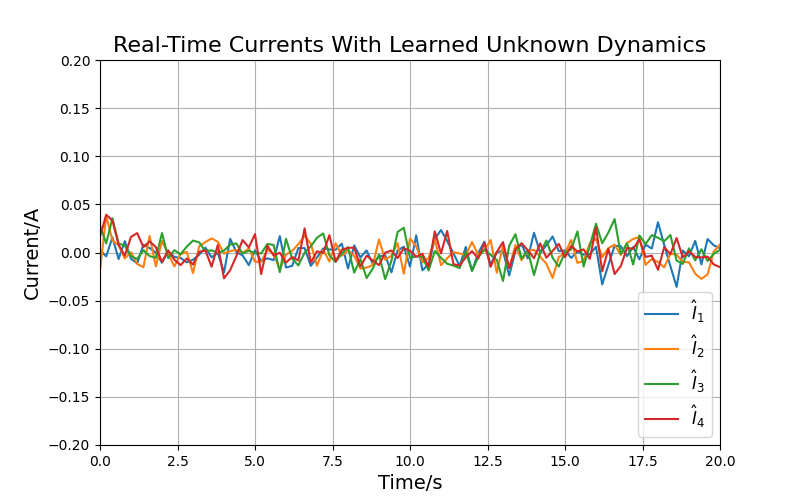}
    \caption{
    Real-time currents for the system under the learned unknown dynamics.}
    \label{fig:real-time_currents_estimated}
    \vspace{-1mm}
\end{figure}

\begin{figure}[!t]
    \centering
    \includegraphics[width=\linewidth]{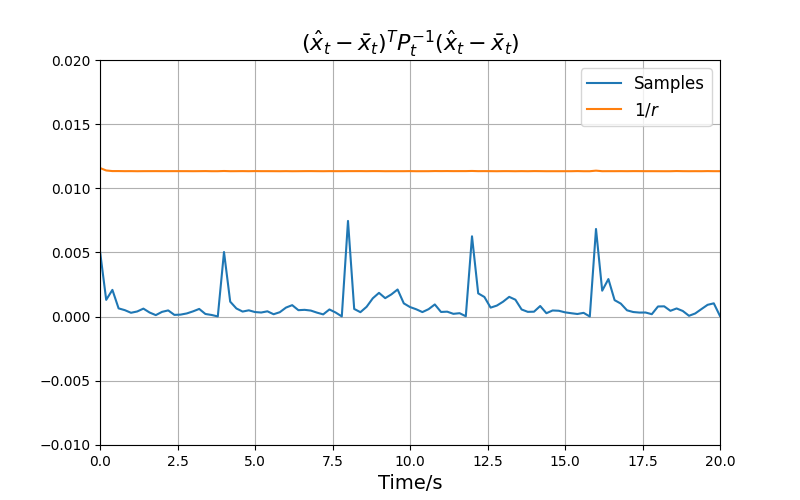}
    \caption{
    Real-time funnel computation for the system under the learned unknown dynamics.}
    \label{fig:real-time_funnel_estimated}
    \vspace{-1mm}
\end{figure}

\begin{figure}[!t]
    \centering
    \includegraphics[width=\linewidth]{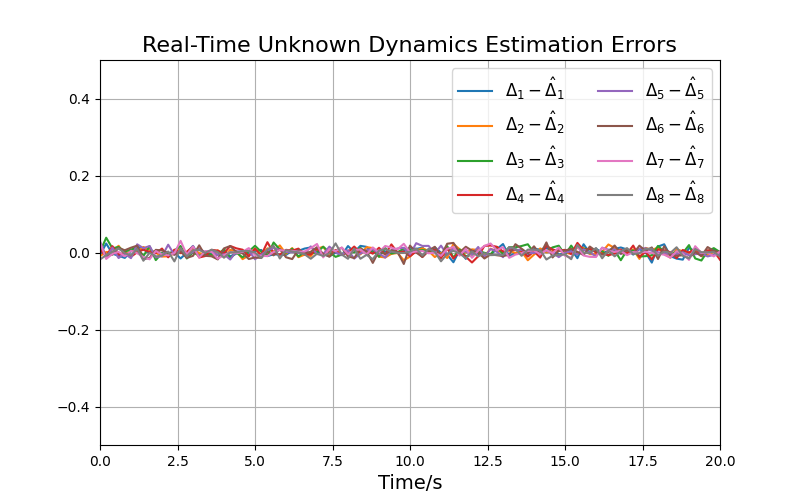}
    \caption{
    Real-time unknown dynamics estimation errors for each elements in $(\Delta_i-\hat{\Delta}_i)$, for all $i\in\mathcal{I}_8$.}
    \label{fig:unknown_dynamics_estimation_errors}
    \vspace{-1mm}
\end{figure}

\begin{figure}[!t]
    \centering
    \includegraphics[width=\linewidth]{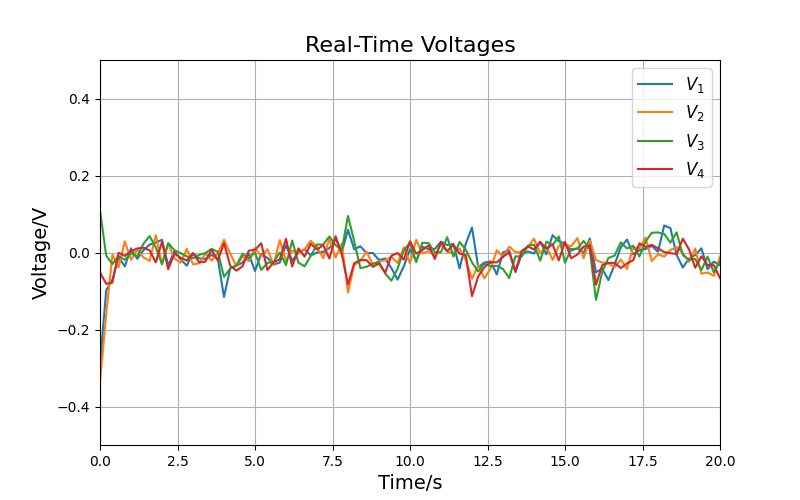}
    \vspace{-2mm}
    \caption{
    Real-time voltages for the actual system.}
    \label{fig:real-time_voltages}
\end{figure}

\begin{figure}[!t]
    \centering
    \includegraphics[width=\linewidth]{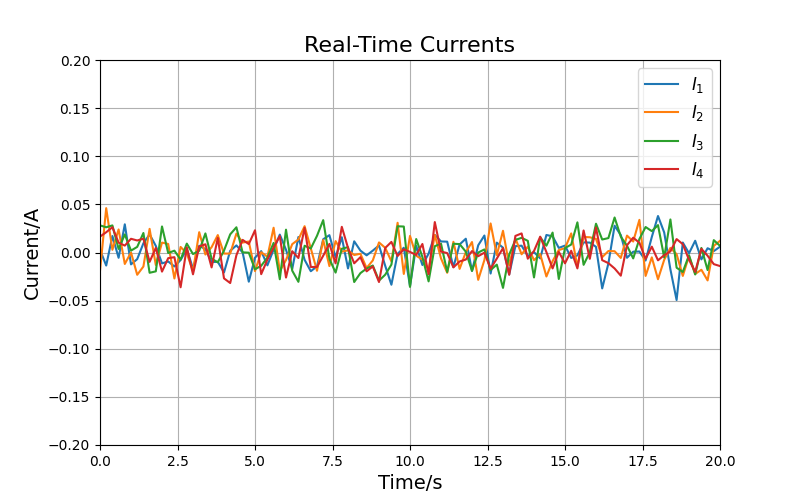}
    \vspace{-2mm}
    \caption{
    Real-time currents for the actual system.}
    \label{fig:real-time_currents}
\end{figure}

\begin{figure}[!t]
    \centering
    \includegraphics[width=\linewidth]{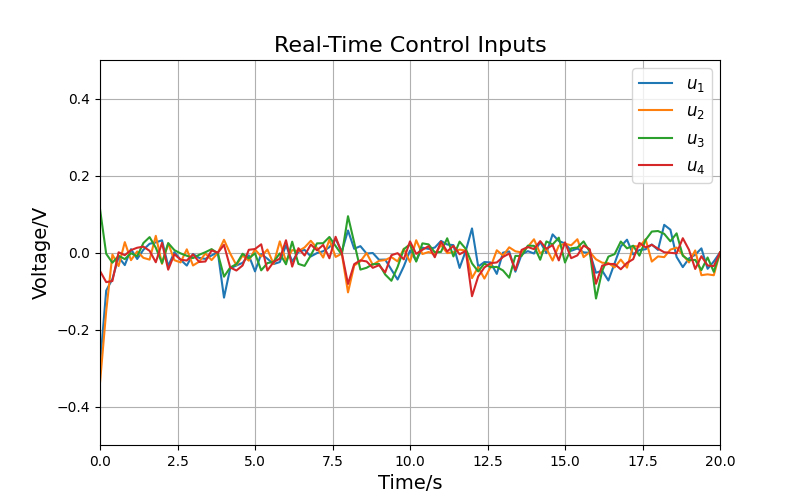}
    \vspace{-2mm}
    \caption{
    Real-time control inputs for the actual system.}
    \label{fig:real-time_control_inputs}
\end{figure}

\begin{figure}[!t]
    \centering
    \includegraphics[width=\linewidth]{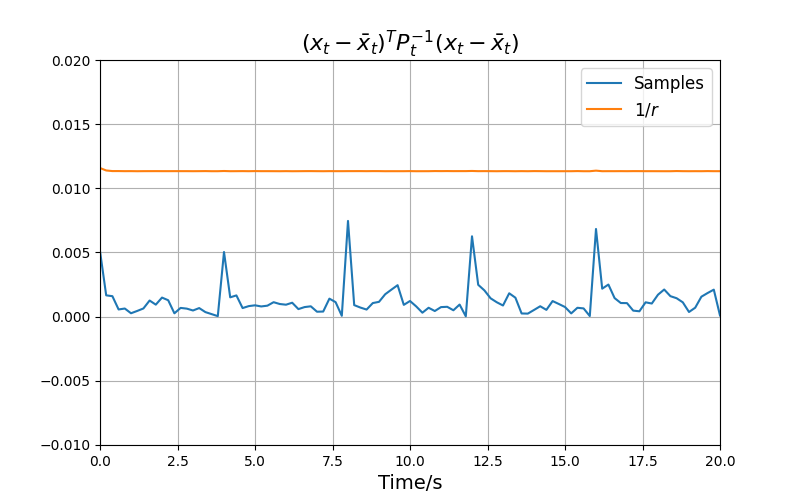}
    \vspace{-2mm}
    \caption{
    Real-time funnel computation for the actual system.}
    \label{fig:real-time_funnel_actual_dynamics}
\end{figure}

The nominal trajectory is conducted through a classical trajectory optimization process, and the results can be seen in Figs. \ref{fig:nominal_voltages}-\ref{fig:nominal_control_inputs}. Here, we only update the nominal trajectory in the first epoch, i.e., $0-4$s, and afterwards, we do not update the nominal trajectory data in case to mimic the attack on the voltage channel. Every $4$ seconds is a training epoch of the REN model (i.e., $5$ training epochs in total), and we reset the initial conditions to add more failures or attacks in the current channel. 
In other words, every epoch is a real-time attacking scenario.
In each epoch, we solve for the funnel controllers and train the REN model, where we only solve the DMI condition for one step and simultaneously train the REN model using the data generated, until we reach the maximum time steps. Therefore, our control synthesis, funnel computation and the training of REN are conducted online using the current states.
Our simulation results can be seen in Figs. \ref{fig:real-time_voltages_estimated}-\ref{fig:real-time_funnel_actual_dynamics}.


It is shown in Figs. \ref{fig:real-time_voltages_estimated}-\ref{fig:real-time_currents_estimated} respectively that the voltages and currents of the system dynamics with the learned unknown dynamics $\hat{\Delta}$ converge to some neighborhood around zero (i.e., the equilibrium points), and the funnel sampling values $(\hat{x}_t-\bar{x}_t)^\T P_t^{-1}(\hat{x}_t-\bar{x}_t)$ converge to some tiny non-negative value close to zero (i.e., $2\times 10^{-5}$) and stay within the estimated funnel size $\frac{1}{r}$ as seen in Fig. \ref{fig:real-time_funnel_estimated}. 
In particular, at the time step $4$s, $8$s, $12$s and $16$s, we reset the initial conditions of the microgrid to generate state deviations caused by faults or attacks. 
Since the control inputs only appear in the voltage channel, it can also be seen in Figs. \ref{fig:real-time_voltages_estimated}-\ref{fig:real-time_currents_estimated} that the voltage deviations are obviously reduced after the deviations (roughly from $0.1$ to $0.014$) while the dynamics of the current are consequently affected, and thus, the control performance of the currents is degraded (just reduce from $0.03$ to $0.01$).
Our results show the robustness and adaptability against the faults and attacks, and thus, the effectiveness of our proposed funnel-based recovery controller is illustrated. 
To show the performance of our REN-based unknown dynamics approximation, we plot the real-time unknown dynamics estimation errors in Fig. \ref{fig:unknown_dynamics_estimation_errors}, where all the components of the approximation errors of the unknown dynamics quickly converge to some small neighborhood around zero (i.e., approximately the range $0\pm 0.015$).

% Real-time states, inputs and funnels for estimated unknown dynamics

The results shown in Figs. \ref{fig:real-time_voltages_estimated}-\ref{fig:unknown_dynamics_estimation_errors} are the ``virtual" dynamics with the learned unknown dynamics, and hence, these results cannot fully illustrate the practical performance of our proposed control framework when applied to the actual dynamics \eqref{Eq:actual_DC_microgrid_subsystem}. Therefore, we show the actual performance by plotting Figs. \ref{fig:real-time_voltages}-\ref{fig:real-time_funnel_actual_dynamics}. Overall, the performance and system behaviors of the actual dynamics under our proposed control framework are similar to those in Figs. \ref{fig:real-time_voltages_estimated}-\ref{fig:real-time_funnel_estimated}. The major differences lie in the enlargement of the magnitude of the actual current values at the equilibrium points, which increases from $0.01$ to $0.02$. 
This is probably because we design our funnel-based controllers for the ``virtual" dynamics with the learned unknown dynamics $\hat{\Delta}$ rather than the actual dynamics, since we do not know the values and the properties of the unknown dynamics $\Delta$. This can also be observed from the convergence of the funnel sampling values, where the converged values increase to $3.4\times 10^{-4}$ as compared to the results in Fig. \ref{fig:real-time_funnel_estimated}.

To show the online training ability of our proposed control framework, we count the time cost on the funnel control synthesis and the REN-based. Here, we do not involve the time spent for nominal trajectory optimization, since this part of time is negligible, which is maximally around $0.0017$s. For the main computational operations, the maximal time cost in funnel control synthesis and REN model training are $0.0923$s and $0.0656$s, respectively, and the maximal total time cost is $0.1331$s, which is lower than the sampling period $0.2$s. 
Therefore, this shows the ability of online implementation of our proposed method.
Moreover, we believe that these solving time can be further reduced if more efficient PCs are applied in practice, and better accuracy can be achieved.

%-------------------------------------------------------------
\section{Conclusion}\label{sec:conclusion}

In this paper, we proposed a funnel-based control method for system recovery with unknown dynamics. By transforming the system dynamics into the Lur'e type dynamics, the Recurrent Equilibrium Network (REN) was applied to learn the unknown dynamics caused by attacks or failures. Considering the system dynamics with this learned unknown dynamics, we designed the funnel-based recovery controller and the internal and $L_2$ stability of the closed-loop system were proved. Furthermore, the invariant funnels were derived, where the states were ensured to stay within the funnels along the nominal trajectory. Moreover, we provided the online implementation of our proposed funnel-based controller. Simulation results showed the effectiveness of our proposed method.
Future work may consider decentralized funnel-based control design and the learning of REN model, so that the scalability can be improved.

% Real-time states and inputs

% %---------------------------------------------------------------
% \section{Appendix}

% \begin{proof}
    
% \end{proof}

%---------------------------------------------------------------
\bibliographystyle{IEEEtran}
\bibliography{references}

\end{document}